\documentclass[journal]{IEEEtran}
 
\usepackage{stmaryrd}
\usepackage{amsmath}
\usepackage{amsthm}
\usepackage{xfrac}
\usepackage{amssymb}
\usepackage{mathabx}
\usepackage{mathrsfs}
\usepackage{psfrag}
\usepackage{graphicx}
\usepackage{epstopdf}
\usepackage{todonotes}
\usepackage{cite}
\usepackage{algorithmic}
\usepackage{algorithm}
\usepackage{svg}
\usepackage{mathtools}
\usepackage{hyperref}
\usepackage{hyphenat}
\usepackage{booktabs}
\usepackage{accents}

\newlength{\dhatheight}

\DeclareMathOperator*{\argmin}{arg\,min}

\DeclareMathOperator{\spanset}{span}

\DeclareMathOperator{\diag}{diag}

\usepackage{float}
\usepackage{multirow}
\usepackage{xparse,mathtools}

\newtheorem{thm}{Theorem}
\newtheorem{corollary}{Corollary}[thm]
\newtheorem{lemma}{Lemma}

\newtheorem{proposition}{Proposition}

\newcommand{\mtrian}{\mathrel{\raisebox{-0.1ex}{%
\scalebox{0.8}[0.6]{$\vartriangle$}}}}

\newcommand\Alpha{\mathrm{A}}
\newcommand\Beta{\mathrm{B}}

\begin{document}
\title{{Universal Approximation of Linear Time-Invariant (LTI) Systems through RNNs: Power of Randomness in Reservoir Computing}}
\author{Shashank Jere, Lizhong Zheng, Karim Said,  and Lingjia Liu
\thanks{S. Jere, K. Said and L. Liu are with Wireless@Virginia Tech, Bradley Department of ECE at Virginia Tech. L. Zheng is with EECS Department at Massachussets Institute of Technology. This work was supported in part by U.S. National Science Foundation (NSF) through grants NSF/CNS-2003059. The corresponding author is L. Liu (ljliu@ieee.org).}
}

\maketitle

\begin{abstract}
Recurrent neural networks (RNNs) are known to be universal approximators of dynamic systems under fairly mild and general assumptions.
However, RNNs usually suffer from the issues of vanishing and exploding gradients in standard RNN training.
Reservoir computing (RC), a special RNN where the recurrent weights are randomized and left untrained, has been introduced to overcome these issues and has demonstrated superior empirical performance especially in scenarios where training samples are extremely limited.
On the other hand, the theoretical grounding to support this observed performance has yet been fully developed.
In this work, we show that RC can universally approximate a general linear time-invariant (LTI) system. 
Specifically, we present a clear signal processing interpretation of RC and utilize this understanding in the problem of approximating a generic LTI system. 
Under this setup, we analytically characterize the optimum probability density function for configuring (instead of training and/or randomly generating) the recurrent weights of the underlying RNN of the RC.
Extensive numerical evaluations are provided to validate the optimality of the derived distribution for configuring the recurrent weights of the RC to approximate a general LTI system.
Our work results in clear signal processing-based model interpretability of RC and provides theoretical explanation/justification for the power of randomness in randomly generating instead of training RC's recurrent weights.
Furthermore, it provides a complete optimum analytical characterization for configuring the untrained recurrent weights, marking an important step towards explainable machine learning (XML) to incorporate domain knowledge for efficient learning.

\end{abstract}


\section{Introduction}
The rise of deep learning methods~\cite{Goodfellow-et-al-2016} in recent times has been unprecedented, owing largely to their remarkable success in fields as diverse as image classification~\cite{HeImageRecognition2016}, speech recognition~\cite{GravesSpeechRecognition2013} and language translation~\cite{bahdanau2014neural}, among many others.
Specifically, recurrent neural networks (RNNs) are known to be universal approximators for dynamical systems under general conditions~\cite{FUNAHASHI1993801}, making them suitable for applications involving temporally correlated data.
Therefore, RNNs are well suited to sequential tasks such as sentence sentiment classification~\cite{AGUEROTORALES2021107373}, language translation~\cite{cho-etal-2014-learning}, video frame analysis~\cite{Guera2018,Zhao2019} as well as recently in receive processing tasks such as symbol detection~\cite{mosleh2017brain} in wireless communications.
More recently, RNNs have also been adapted to be applied in natural language processing (NLP) tasks to emulate the remarkable success of transformers~\cite{peng2023rwkv} while avoiding their high computational and memory complexity.
However, vanilla RNNs exhibit the problem of vanishing and exploding gradients~\cite{pascanu2013difficulty} when trained using the backpropagation through time (BPTT) algorithm~\cite{Werbos1990}.
Long short-term memory (LSTM) networks~\cite{Hochreiter1997a} alleviate this problem to a certain degree by incorporating additional internal gating procedures~\cite{Bengio1994, Hochreiter1996} and thus, deliver more robust performance compared to vanilla RNNs~\cite{SHERSTINSKY2020}.
On the other hand, LSTMs require significantly more training data due to their richer modeling capabilities, thereby posing a challenge when the training data is inherently limited, e.g., in the physical (PHY) and medium access control (MAC) layers of modern wireless systems where the over-the-air (OTA) training data is extremely limited.
To balance this trade-off, randomized recurrent neural networks~\cite{Gallicchio2018RandomizedRN} have been a topic of active investigation. 
A general randomized RNN consists of an untrained hidden layer with recurrent units, which non-linearly projects the input data into a high-dimensional feature space, and a trained output layer which scales and combines the outputs of the hidden layer in a linear fashion.
Reservoir Computing (RC)~\cite{lukovsevivcius2009reservoir} is a specific paradigm within the class of randomized RNN approaches where the echo state network (ESN)~\cite{Lukosevicius2012} is a popular implementation of the general RC framework.

In RC architectures including the ESN, typically only the output layer of the network is trained using pseudo-inversion or Tikhonov regularization, while the weights of the input layers and the hidden layers are fixed after initialization based on a certain pre-determined distribution.
This particular feature of RC significantly reduces the amount of required training making it uniquely suitable for applications where the number of training samples is extremely limited.
Furthermore, since the recurrent weights are randomly generated and fixed, RC completely avoids the issues of vanishing and exploding gradients that commonly occur in the standard RNN training.
Despite its limited training, RC has demonstrated impressive performance in many sequential processing applications including NLP tasks, e.g., decoding grammatical structure from sentences~\cite{Hinaut2012}, learning word-to-meaning mappings~\cite{Juven2020}, in video frame analysis tasks such as event detection in visual content~\cite{Jalalvand2015}, as well as in stock market prediction~\cite{WANG2021115022}.
Recently, RC has found great appeal in various wireless applications, especially in the PHY/MAC layer receive processing with extremely scarce OTA training data.
For example, ESNs and its extensions have been utilized to construct symbol detectors for 5G and Beyond 5G multiple antenna systems~\cite{mosleh2017brain,zhou2019,zhou2020rcnet,RCStruct_TWC}.
In addition, the ESN has been applied to effectively combat inter-symbol interference (ISI) and improve detection performance in a chaotic baseband wireless communication system~\cite{Ren2020}.
Furthermore, ESN-based deep reinforcement learning has been introduced for dynamic spectrum access in 5G networks to provide improved sample efficiency and convergence rate over traditional RNN structures~\cite{HaoHsuanDEQN2022}.
Beyond conventional wireless communications, RC has also found utility in equalization for optical transmission~\cite{DaRos2020} and signal classification in optoelectronic oscillators~\cite{Dai2021}. 

Although RNNs and its variants including RC have shown superior empirical performance in various sequence processing tasks, a fundamental theoretical understanding of their effectiveness using classical tools remains largely unexplored. 
As discussed in~\cite{Rubayet20AI}, ``lack of explainability'' is one of the top five challenges in applying machine learning approaches to applications with limited training data such as telecommunication networks, which have traditionally been designed based on a mixture of theoretical analysis, wireless channel measurements, and human intuition and understanding. 
In fact, the traditional approach has proven amenable for domain experts to resort to either theoretical analysis or computer simulations to validate wireless system building blocks. 
Therefore, it is desirable for neural network models to have similar levels of explainability especially when designed for wireless systems, and in general for applications with specifications-limited or cost-prohibitive procedures of obtaining training data samples.

\subsection{State-of-the-Art in Explainable Machine Learning (XML)}
Even though deep neural networks have been effective in various applications, they are still largely perceived as black-box functions converting features in input data to classification labels or regression values at their output.
With the growing real-world application of neural network models in sensitive areas such as autonomous driving and medical diagnostics, there is an increasing need to develop a deep understanding of the inner workings of such models. 
This has given birth to the field of Explainable Machine Learning (XML) which has seen important developments in recent times.
A useful overview of Layer-Wise Relevance Propagation (LRP), which is an explainability technique for deep neural networks that uses propagation of relevance information from the output to the input layers, is provided in~\cite{Montavon2019}.
An information-theoretic approach towards opening the black box of neural networks was provided in~\cite{shwartzziv2017opening} building upon the information bottleneck (IB) principle.
SHAP (SHapley Additive exPlanations), which is a model interpretation framework built on the principles of game theory, was introduced in~\cite{Lundberg2017}.
Outside of neural network models, the work in~\cite{Baehrens2010} introduces the concept of local explanation vectors, applying the technique to support vector machines (SVMs). 
While these works introduce useful interpretation and explanation frameworks, a first principles-based approach that utilizes a signal processing-oriented understanding is largely missing or not yet fully developed for most neural network architectures.

Among studies exploring the theoretical explanations behind the success of RC in time-series problems, one of the first is~\cite{Ozturk2007}, which introduces a functional space approximation framework for a better understanding of the operation of ESNs.
Another recent work of note is~\cite{Bollt2021} which shows that an ESN without nonlinear activation is equivalent to vector autoregression (VAR). 
\cite{HART2021} makes the case for ESNs being universal approximators for ergodic dynamical systems.
The effectiveness of RC in predicting complex nonlinear dynamical systems such as the Lorenz and the Rössler systems was studied in~\cite{Halus2019}, while~\cite{Carroll2022} investigated the tuning and optimization of the length of the fading memory of RC systems. 
Our previous work in~\cite{JereTCOM2023} derived an upper bound on the Empirical Rademacher Complexity (ERC) for single-reservoir ESNs and showed tighter  generalization for ESNs as compared to traditional RNNs, while simultaneously demonstrating the utility of the derived bound in optimizing an ESN-based symbol detector in multi-antenna wireless receivers.
Other statistical learning theory-based works such as~\cite{Gonon2020} also attempt to bound the generalization error for RC using slightly modified Rademacher-type complexity measures.
In our previous work~\cite{Jere2023WCL}, we introduce a signal processing analysis of the ESN and present a complete analytical characterization of the optimum untrained recurrent weight for an ESN with a single neuron when employed in the wireless channel equalization task.
While the works in existing literature provide interesting insights using information-theoretic or statistical learning-theoretic principles, a lucid signal processing understanding coupled with complete analytical characterizations using conventional tools has not been established yet.
With this in mind, we aim to accomplish the following two objectives in this work: i) Gain a theoretical understanding of why randomly generated reservoir weights provide good empirical performance for function approximation, and ii) develop a systematic methodology to configure this random generation of reservoir weights incorporating prior information or domain knowledge.
With these objectives, we provide an outline in the next section for the set of problems considered, the overall approach adopted and the steps taken to solve them.

\subsection{Our Contributions}
The main contributions of this work are summarized below:
\begin{enumerate}    

    \item First, we consider the ``atomic'' problem of approximating the impulse response of a first-order infinite impulse response (IIR) filter using an ESN consisting of two neurons in the reservoir with fixed reservoir weights and with linear activation. Formulating this as an orthogonal projection problem, we precisely calculate the corresponding approximation error and derive an exact scaling law that relates this approximation error to the distance between the ESN's poles (i.e., the recurrent reservoir weights).    

    \item Second, continuing with the impulse response of the first-order IIR system as the target function, we consider the problem of approximating its impulse response using an ESN with multiple neurons having randomly generated weights. Optimizing the corresponding approximation error, we derive the optimum probability density function (PDF) to configure the random generation of the ESN reservoir weights.    

    \item Third, we generalize this result by showing that the derived optimum PDF for approximating a first-order IIR system is also optimum for approximating general higher-order LTI systems that can be written as a linear combination of first-order poles.

    \item Fourth, we show that under linear activation, a reservoir with random and sparse interconnections between its constituent neurons has an equivalent representation as a reservoir with non-interconnected neurons.

    \item Finally, via extensive numerical evaluations, we empirically confirm the following: i) Validity of the derived approximation error scaling law, and ii) Optimality of the derived optimum PDF for configuring the ESN reservoir weights when applied to the task of approximating a first-order IIR and higher-order LTI systems.
    
\end{enumerate}

The rest of the paper is organized as follows. Sec.~\ref{sec:problem_formulation} presents the problem formulation for the task of LTI system approximation using an ESN. Sec.~\ref{sec:reservoir_optimization} presents our approach and analysis to derive the optimum distribution for configuring the random generation of ESN reservoir weights. Sec.~\ref{sec:limited_training} outlines the training procedure of the ESN and briefly outlines overfitting concerns in this scenario. 
Numerical evaluations to validate the theoretical findings in the preceding sections are presented in Sec.~\ref{sec:numerical_evaluation}. 
Finally, we provide concluding remarks and directions for future work in Sec.~\ref{sec:conclusion}. 

\textbf{Notation:}
$\mathbb{R}$: set of real numbers; $\mathcal{U}(a,b)$: uniform distribution with support $[a,b]$; $\mathcal{N}(\mu, \sigma^2)$: Gaussian (normal) distribution with mean $\mu$ and variance $\sigma^2$; 
$\mathbb{E}[\cdot]$: Expectation operator, $\mathsf{VAR}(\cdot)$: Variance operator;
$c$ and $C$ denote scalars, 
$\mathbf{c}$ denotes a column vector;
$\| \cdot \|_2$: $\ell_2$-norm; 
$\langle \mathbf{a}, \mathbf{b} \rangle = \mathbf{b}^T \mathbf{a}$: inner product of vectors $\mathbf{a}, \mathbf{b} \in \mathbb{R}^n$.
$\mathbf{C}$ denotes a matrix;
$(\cdot)^T$: matrix transpose; $(\cdot)^{\dagger}$: Moore-Penrose matrix pseudo-inverse.  
$\lambda(\mathbf{C})$ denotes the spectrum (set of eigenvalues) of $\mathbf{C}$.
$p_{\Alpha}(\cdot)$ denotes the probability density function (PDF) of a random variable $\alpha$. 
$\Pr(E)$ denotes the probability of event $E$.
$(a, b)$ denotes an open interval and $[a, b]$ denotes a closed interval for $a, b \in \mathbb{R}$.
W.L.O.G. stands for ``without loss of generality''.
We define the following terms to have this specific meaning in the remainder of the paper: i) ``Training'': Data-driven optimization of neural network (NN) model weights via backpropagation-based or single-shot algorithms (e.g., least squares), ii) ``Randomly generating'': The process of generating NN model weights in an i.i.d. manner by sampling them from a pre-determined and unoptimized distribution, iii) ``Configuring'': The process of generating NN model weights in an i.i.d. manner by sampling them from an analytically derived distribution taking into account prior information or domain knowledge.

\section{Problem Formulation}
\label{sec:problem_formulation}

\subsection{Randomized RNN: RC and The Echo State Network (ESN)}
\label{sec:conventional_esn}
In the context of a randomized RNN~\cite{Gallicchio2018RandomizedRN} and more specifically an ESN, a general learning problem can be defined by the tuple $(\mathcal{Z}, \mathcal{P}, \mathcal{H}, \ell)$, where:
\begin{itemize}
\item $\mathcal{X}$ and $\mathcal{Y}$ are the input and output spaces respectively. In our case, $\mathcal{X} \in \mathbb{R}^{D \times T}$ represents a time sequence of length $T$. The output space is $\mathcal{Y} \in \mathbb{R}^{K \times T}$ or $\mathcal{Y} \in \{0,1\}^{K \times T}$, depending on whether the network is being employed for regression or classification respectively in the sequence-to-sequence setting. In the sequence-to-vector setting, we have $\mathcal{Y} \in \mathbb{R}^{K}$ or $\mathcal{Y} \in \{0,1\}^{K}$. Here, $D$ and $K$ are the input and output dimensions respectively.

\item  $\mathcal{Z} = \mathcal{X} \times \mathcal{Y}$ represents the joint input-output space. 

\item $\mathcal{P}$ is the space of probability distributions defined on $\mathcal{Z}$. 

\item $\mathcal{H}$ is the space of all function approximators $h:\mathcal{X} \rightarrow \mathcal{Y}$ where $h$ denotes the neural network function. 

\item The loss function $\ell(\cdot)$ is defined as 
$\ell: \mathcal{Y} \times \mathcal{Y} \rightarrow \mathbb{R}$.

\end{itemize}

\begin{figure}[htbp]
    \centering    \includegraphics[width=0.9\linewidth]{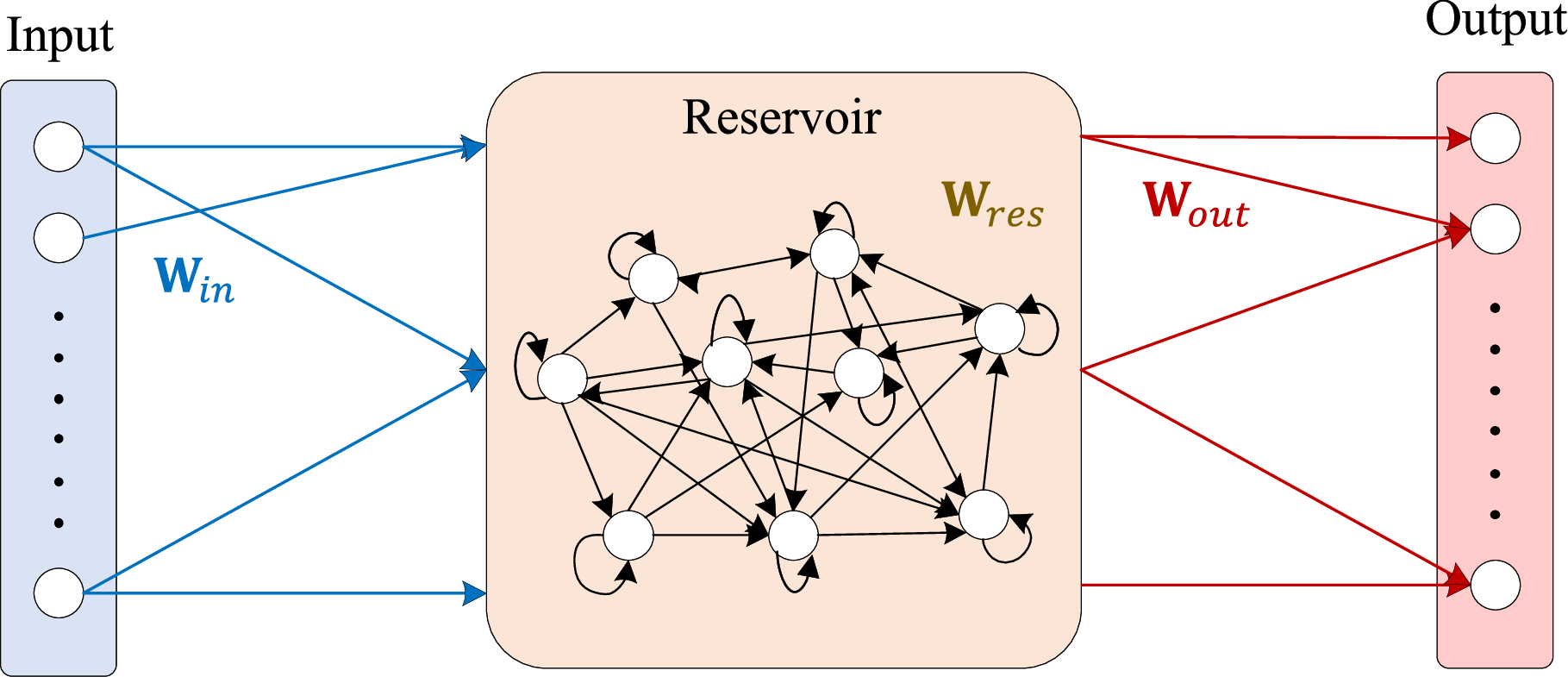}
    \caption{An echo state network (ESN) with a single reservoir.}
    \label{fig:ESN_figure}
\end{figure}

Define an input sequence $\mathbf{U} = \big[\mathbf{u}[1], \mathbf{u}[2], \cdots, \mathbf{u}[T]\big]$ of length $T$ such that $\mathbf{u}[n] \in \mathbb{R}^D$ and $\mathbf{U} \in \mathbb{R}^{D \times T}$ for the discrete-time indices $n = 1,2,\cdots,T$. Note that each data sample $\mathbf{u}(n)$ in the time series $\mathbf{U}$ is a (column) vector of dimension $D$. 
For every training sequence $\mathbf{U}$, a label (ground truth) sequence $\mathbf{G}$ is available to train the network, where $\mathbf{G} = \left[\mathbf{g}[1], \mathbf{g}[2], \cdots, \mathbf{g}[T] \right]$ such that $\mathbf{g}[n] \in \mathbb{R}^K$ for a (sequence-to-sequence) regression task and $\mathbf{g}[n] \in \{0,1\}^K$ for a (sequence-to-sequence) classification task. 
The training set $\mathsf{Z}^N$ of size $N$ is then defined as the set of input-label tuples $\mathsf{Z}^N := \big\{(\mathbf{U}_1, \mathbf{G}_1), (\mathbf{U}_2, \mathbf{G}_2), \cdots, (\mathbf{U}_N, \mathbf{G}_N) \big\}$,
where $\mathsf{Z}^N$ is generated i.i.d. according to some (unknown) joint input-output probability distribution $\mathsf{P}(\cdot, \cdot) \in \mathcal{P}$.
The general setup described above is applicable to a time series problem with any recurrent deep learning model.
Within the class of randomized RNNs, we consider a single reservoir ESN containing $M$ neurons with random and sparse interconnections  (among other possibilities) and a single output (readout) weights matrix. This structure is depicted in Fig.~\ref{fig:ESN_figure}.
Next, we define the input, output and the model weights of the ESN in the following: 
\begin{itemize}
    
    \item $\mathbf{x}_{\mathrm{res}}[n] \in \mathbb{R}^{M}$ is the reservoir state vector at time index $n$.
    
    \item $\mathbf{X}_{\mathrm{res}} = \big[\mathbf{x}_{\mathrm{res}}[1], \cdots, \mathbf{x}_{\mathrm{res}}[T] \big] \in \mathbb{R}^{M \times T}$ is defined as the ``reservoir states matrix'' of the individual states from $n=1$ to $n=T$.

    \item $\mathbf{x}_{\mathrm{in}}[n] \in \mathbb{R}^{D}$ denotes the ESN input.   
    $\mathbf{y}[n] \in \mathbb{R}^{K}$ denotes the ESN output.
    
    \item $\mathbf{W}_{\mathrm{in}} \in \mathbb{R}^{M \times D}$ is the input weights matrix, $\mathbf{W}_{\mathrm{res}} \in \mathbb{R}^{M \times M}$ is the reservoir weights matrix, $\mathbf{W}_{\mathrm{out}} \in \mathbb{R}^{K \times M}$ is the output weights matrix. 
    
\end{itemize}
For a point-wise nonlinear activation function $\sigma(\cdot)$, the state update equation and the output equation are respectively:
\begin{align}
\mathbf{x}_{\text{res}}[n] &= \sigma \big(\mathbf{W}_{\text{res}}\mathbf{x}_{\text{res}}[n-1] + \mathbf{W}_{\text{in}}\mathbf{x}_{\text{in}}[n] \big) \label{eq:esn_state_eqn}, \\
\mathbf{y}[n] &= \mathbf{W}_{\text{out}}\mathbf{x}_{\text{res}}[n].
\label{eq:output_eqn}
\end{align}
In this setup, $\mathbf{W}_{\text{in}}$ and $\mathbf{W}_{\text{res}}$ are randomly generated, i.e., initialized from a certain pre-determined but arbitrary distribution, e.g., the uniform or Gaussian distributions, and then kept fixed throughout the training and inference (test) stages. 
Unlike vanilla RNNs and its variants where all network weights are trained using BPTT, the only trainable network parameter in the ESN is the output weights matrix $\mathbf{W}_{\text{out}}$, which is trained using a pseudoinverse-based closed-form linear update rule.
This greatly reduces the number of trainable parameters, as well as the training computational complexity, lending well to applications with limited training data availability.
Additionally, the sparsely interconnected nature of $\mathbf{W}_{\text{res}}$ is controlled via the hyperparameter named `sparsity' (denoted as $\kappa$), which represents the probability of an element of $\mathbf{W}_{\text{res}}$ being zero. The internal reservoir structure of Fig.~\ref{fig:ESN_figure} depicts this random and potentially sparse nature of the interconnections between the constituent neurons.

\subsection{Approximating an Atomic LTI System with an ESN}
\label{sec:lti_approximation}
Consider the target LTI system characterized by the following causal time-domain impulse response:
\begin{align}
    \mathbf{s}_{\alpha}[n] = &\begin{cases}
        \alpha^n, & n \geq 0 \\
        0, & n < 0
    \end{cases} 
    = \alpha^n u[n],
    \label{eq:exponential}
\end{align}
where $\alpha \in (-1, 1)$ and $u[n]$ is the discrete-time unit step function. 
Thus, the target system to be approximated by the ESN is described by the time-domain impulse response characterized by the infinite-dimensional vector $\mathbf{s}_{\alpha} \in \mathbb{R}^{\infty}$. 
We choose this as the first case to analyze since the time-domain impulse response of a large class of general LTI systems can be written exactly as a linear combination of first order IIR impulse responses of Eq.~\eqref{eq:exponential}~\cite{Oppenheim1996}, i.e., 
\begin{align}
    \mathbf{h}[n] = \sum_{i=1}^{N_0} w_i  \mathbf{s}_{\alpha_i}[n],
\end{align}
where $w_i \in \mathbb R$ are the combining weights, thereby making the extension to the general case feasible given the analysis for the simple case of Eq.~\eqref{eq:exponential}. 
This is shown in Sec.~\ref{sec:extension_to_higher_order_lti}.

\begin{figure}[htbp]
    \centering    \includegraphics[width=0.85\linewidth]{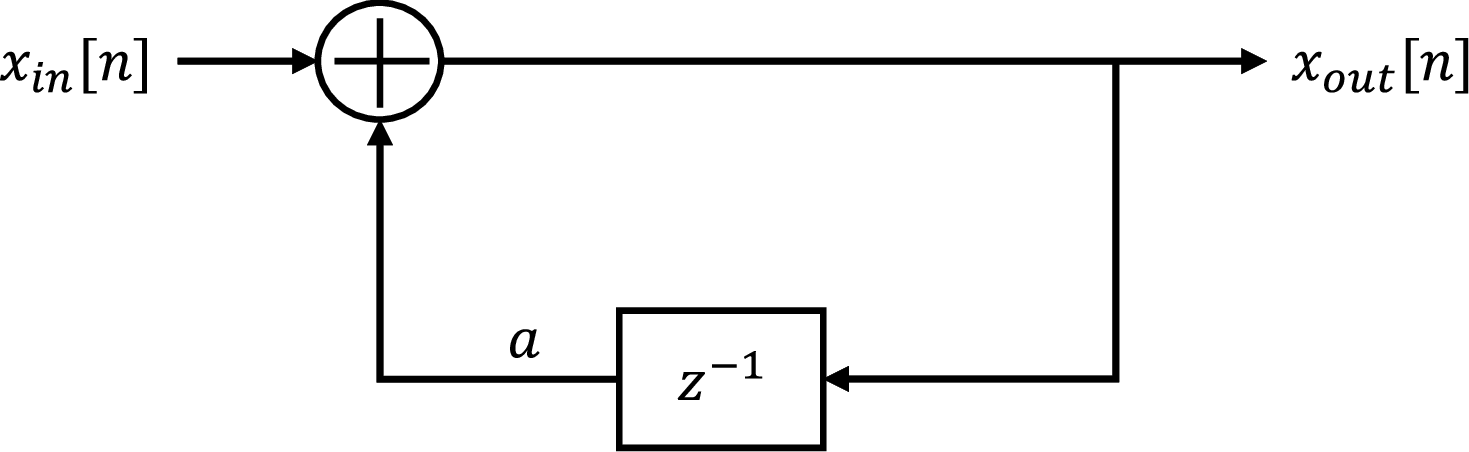}
    \caption{Modeling a neuron in the reservoir with linear activation as a single-pole IIR filter.}
    \label{fig:single_pole_iir}
\end{figure}

In this work, we consider a simplified version of the more general ESN described in Sec.~\ref{sec:conventional_esn}. Specifically, we consider a simple reservoir construction where the individual neurons are disconnected from each other and only consist of unit delay self-feedback loops. 
This translates to $\mathbf{W}_{\mathrm{res}}$ being a diagonal matrix.
Next, for tractability of analysis, we consider linear activation such that $\sigma(\cdot)$ in Eq.~\eqref{eq:esn_state_eqn} is an identity mapping.
As shown in our previous work~\cite{Jere2023WCL}, a single neuron with linear activation can be modeled as a first-order infinite impulse response (IIR) filter with a single pole. This is illustrated in Fig.~\ref{fig:single_pole_iir}, where a single neuron or ``node'' inside the reservoir simply implements a first-order autoregressive process AR(1) with a feedback weight $a$.
The system response for the IIR filter in Fig.~\ref{fig:single_pole_iir} is given by 
\begin{align}
    H_{0}(z) = \frac{X_{\mathrm{out}}(z)}{X_{\mathrm{in}}(z)} = \frac{1}{1 - az^{-1}}.
\end{align}
Finally, we consider the input weights to be unity, as their effect is absorbed in the output weights when the activation employed in the reservoir is linear.

\begin{figure}[htbp]
    \centering    \includegraphics[width=\linewidth]{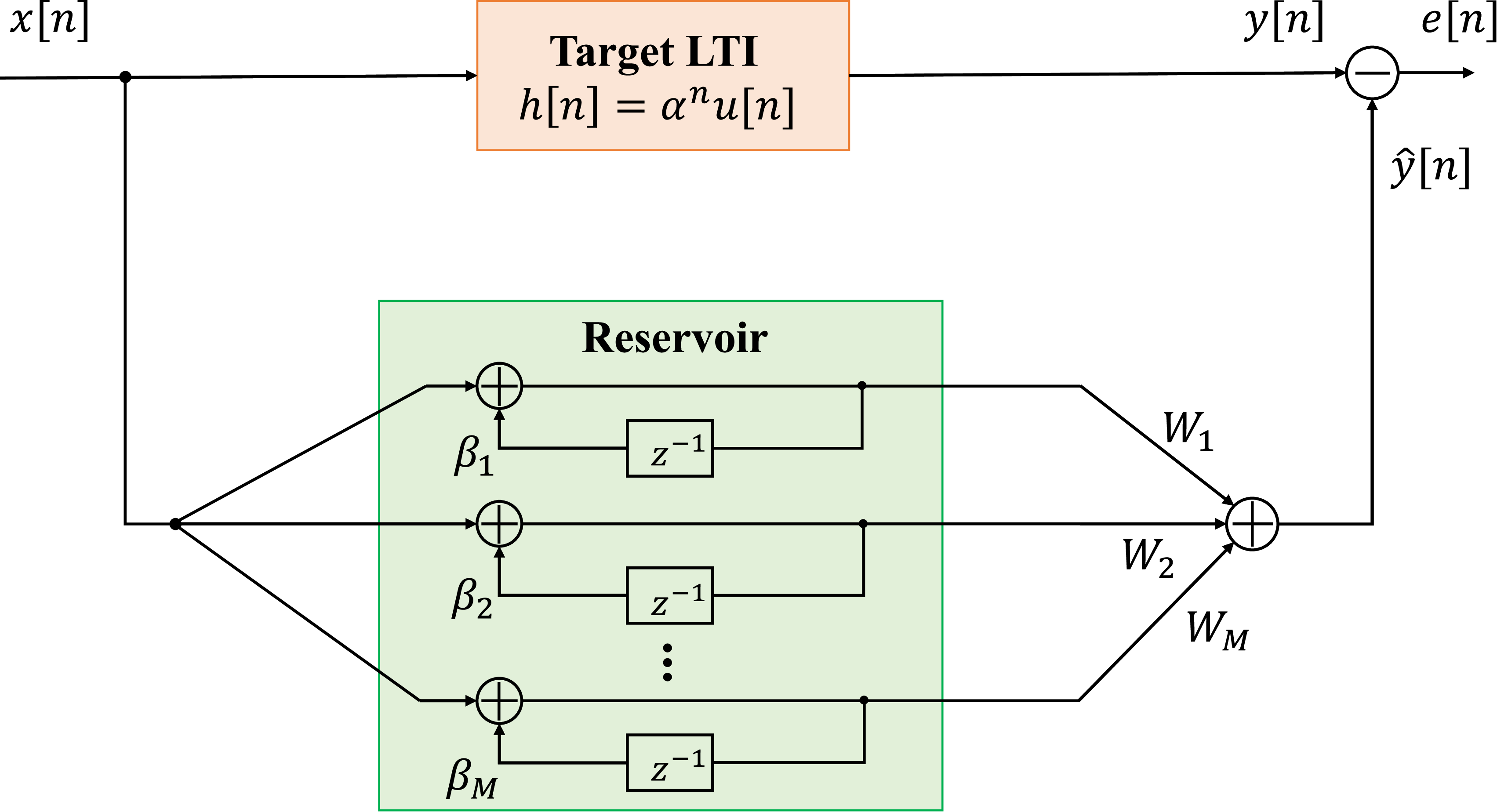}
    \caption{Approximating an LTI system belonging to a known model family (e.g., first-order IIR system) with a linear non-interconnected reservoir ESN.}
    \label{fig:LTI_system_simulation}
\end{figure}

With the aforementioned preliminaries laid out, the ESN design problem for LTI system approximation can be articulated as follows. 
Consider an ESN reservoir as a collection of non-interconnected neurons with fixed corresponding reservoir (recurrent) weights $\{\beta_m\}_{m=1}^{M}$, where each $\beta_m \in (-1, 1)$ to ensure stability of its impulse response $\beta_m^n u[n]$.  
Such a reservoir with non-interconnected neurons has also been considered for neuromorphic computing in an experimental setting using photonic hardware~\cite{Sozos2022}, thus highlighting its practical applicability.
We would like to choose $\{\beta_m \}_{m=1}^{M}$ such that their weighted combination approximates the normalized target $\frac{\mathbf{s}_{\alpha}}{\|\mathbf{s}_{\alpha} \|_{2}}$ with a low approximation error, i.e., 
\begin{align}
    \frac{\mathbf{s}_{\alpha}}{\|\mathbf{s}_{\alpha} \|_{2}} \approx \sum_{m=1}^{M} W_m \mathbf{s}_{\beta_m},
    \label{eq:projection_problem}
\end{align}
where $\mathbf{s}_{\beta_m}[n] = \beta_m u[n]$.
Note that target normalization is imperative to ensure that the mean approximation error across multiple LTI system realizations (i.e., values of ``parameter'' $\alpha$) is not dominated by realizations for which $\alpha$ is closer to $1$ or $-1$ over those for which $\alpha$ is closer to $0$. With this formulation, the normalized target has unit norm.
This can also be written as the system function in terms of the $z$-transform as
\begin{align}
    S_{\alpha}(z) \approx \sum_{m=1}^{M} W_m S_{\beta_m}(z),
\end{align}
which can be expanded as
\begin{align}
    \frac{\sqrt{1 - \alpha^2}}{1-\alpha z^{-1}} \approx \sum_{m=1}^{M} \frac{W_m}{1 - \beta_m z^{-1}}.
\end{align}
Thus, the system being estimated is an infinite impulse response (IIR) system with a single pole $\alpha$, where the ESN attempts to estimate this IIR impulse response as a weighted combination of $M$ IIR impulse responses characterized by the random poles $\{\beta_m \}$, which are kept fixed during training of the output weights $\{W_m \}$ and during test.
This problem can be characterized as a system ``approximation'' or ``identification'' problem, whereby a linear ESN with a reservoir of non-interconnected neurons with randomly generated or configured weights attempts to reproduce the output of the unknown LTI system belonging to a known model family, in this case, a single-pole IIR system. 
The problem setup is depicted in Fig.~\ref{fig:LTI_system_simulation}.

\subsection{ESN Initialization and Training}
\begin{figure}[htbp]
    \centering    \includegraphics[width=0.775\linewidth]{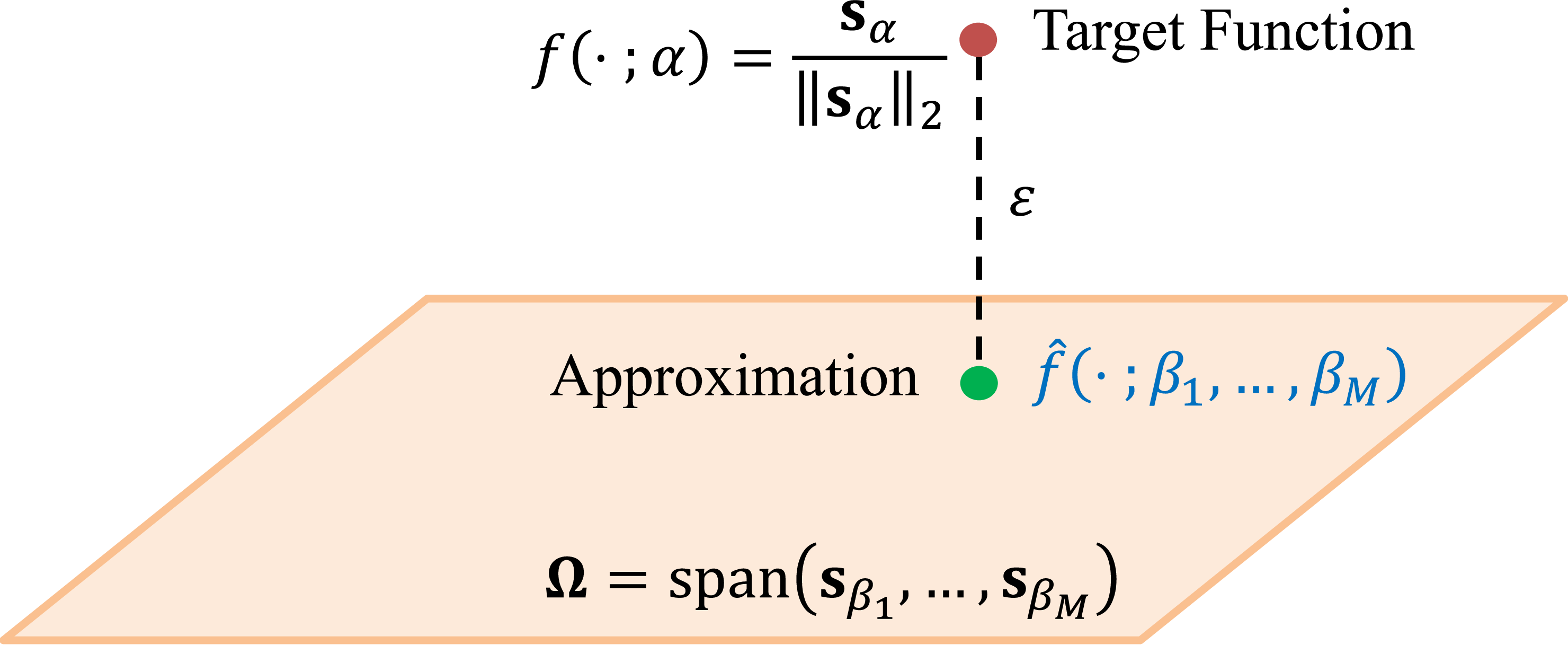}
    \caption{Learning a single-pole IIR system: An orthogonal projection view.}     
    \label{fig:projection_geometry}
\end{figure}

The process of initializing the ESN reservoir weights (random generation or configuration) and subsequent training of output weights consists of three components:
i) a target function $f(\cdot; \alpha)$ to be approximated, ii) a linear subspace $\boldsymbol \Omega$ spanned by the reservoir basis functions, and iii) an approximation $\widehat{f}(\cdot; \beta_1, \ldots, \beta_M)$ of the target function in $\boldsymbol \Omega$.
For the LTI system approximation task, the target function is the normalized impulse response of the system, given by $f(\cdot; \alpha) = \frac{\mathbf{s}_{\alpha}}{\|\mathbf{s}_{\alpha} \|_{2}}$.
The subspace spanned by the reservoir neurons is given by $\boldsymbol{\Omega} = \spanset(\mathbf{s}_{\beta_1}, \ldots, \mathbf{s}_{\beta_M})$.
For a general loss function $\mathcal{L}(f;\widehat{f})$, the training procedure finds the output combining weights $\{W_m \}$ such that the approximation $\widehat{f}(\cdot; \beta_1, \ldots, \beta_M)$ lying in $\boldsymbol\Omega$ minimizes $\mathcal{L}(f;\widehat{f})$.
With the $\ell_{2}$ norm as the loss function $\mathcal{L}(f;\widehat{f})$, the ESN training procedure finds $\widehat{f}(\cdot; \beta_1, \ldots, \beta_M)$ as the $\ell_2$ training loss minimizing approximation, implying that it is the orthogonal projection of $f(\cdot; \alpha)$ onto $\boldsymbol{\Omega}$.
The corresponding approximation error is then referred to as the ``projection error''.
This setup is illustrated in Fig.~\ref{fig:projection_geometry}.
The projection error can be written as the following $\ell_2$-loss: 
\begin{align}
    \varepsilon = \left\| \frac{\mathbf{s}_{\alpha}}{\|\mathbf{s}_{\alpha} \|_{2}} - \sum_{m=1}^{M} W_m^* \mathbf{s}_{\beta_m} \right\|_{2}^2,
    \label{eq:training_l2_loss}
\end{align}
where
\begin{align}
    \{ W_m^* \} = \argmin_{ \{W_m \} } \left\| \frac{\mathbf{s}_{\alpha}}{\|\mathbf{s}_{\alpha} \|_{2}} - \sum_{m=1}^{M} W_m \mathbf{s}_{\beta_m} \right\|_{2}^{2},
\end{align}
are the optimum output weights given by 
\begin{align}
    \mathbf{w} = \mathbf{\Sigma}^{-1} \mathbf{r}.
    \label{eq:optimum_output_weights}
\end{align}
Here, $\mathbf{w} \overset{\mtrian}{=} [W_1^* \; \ldots \; W_M^*]^T \in \mathbb{R}^{M}$, and the projection error can be shown to be
\begin{align}
    \varepsilon = 1 - \mathbf{r}^T \mathbf{\Sigma}^{-1} \mathbf{r}.
    \label{eq:simplified_projection_error}
\end{align}
Here, $\|\mathbf{s}_{\alpha} \|_{2}^{2} = \sum_{n=0}^{\infty} \alpha^{2n} = \frac{1}{1 - \alpha^2}$, and $\mathbf{r} \in \mathbb{R}^{M \times 1}$ and $\mathbf{\Sigma} \in \mathbb{R}^{M \times M}$ are respectively defined as 
\begin{align}
    \mathbf{r} \overset{\mtrian}{=} \frac{1}{\|\mathbf{s}_{\alpha} \|_2} \begin{bmatrix}
\langle \mathbf{s}_{\beta_1}, \mathbf{s}_{\alpha} \rangle \\
\vdots \\
\langle \mathbf{s}_{\beta_M}, \mathbf{s}_{\alpha} \rangle
\end{bmatrix},
\end{align}
and $[\mathbf{\Sigma}]_{i,j} \overset{\mtrian}{=} \langle \mathbf{s}_{\beta_i}, \mathbf{s}_{\beta_j} \rangle $, where $[\mathbf{\Sigma}]_{i,j}$ is the element of $\mathbf{\Sigma}$ in the $i^{\text{th}}$ row and the $j^{\text{th}}$ column.

The projection error of Eq.~\eqref{eq:simplified_projection_error} is intrinsically linked to the training loss when $\{W_m \}$ are trained with finite labeled data samples. 
Specifically, for an impulse input, i.e., $\|\mathbf{x} \|_2 = 1$, the data-driven training loss is lower bounded by the projection error $\varepsilon$.
This is because the projection error makes use of the ``optimum'' output combining weights $\{W_m^* \}$, computing which requires knowledge of $\alpha$.
This inherently assumes that an infinite number of data samples are available for training $\{W_m\}$ so that they converge to $\{W_m^*\}$.
Thus, the projection error of Eq.~\eqref{eq:simplified_projection_error} is the lowest achievable training loss for an impulse input ($\|\mathbf{x} \| = 1$).
Then, the loss metric $\mathcal{L}(\cdot;\cdot)$ to be used to optimize the fixed reservoir weights $\{\beta_m \}$ can be defined as the projection error of Eq.~\eqref{eq:training_l2_loss}, i.e., 
\begin{align}
    \mathcal{L}(\alpha;\beta_1, \ldots, \beta_M) \overset{\mtrian}{=} \varepsilon.
\end{align}

Since we are interested in designing a single ESN with reservoir weights that provides a low approximation error \emph{on average}, we model $\alpha$ as a random variable with a known prior PDF $p_{\Alpha}(\cdot)$.
For example, system identification tasks in acoustic and electromechanical servo systems employ frequency-domain methods~\cite{LIU2008998, ZHANG2005367} in practice to empirically deduce the distribution of the system poles or the modes of a given LTI system.
Then, the ultimate ESN design goal is to analytically choose the fixed optimum reservoir weights $\{ \beta_1^*, \ldots, \beta_M^* \}$ according to
\begin{align}
    \{ \beta_1^*, \ldots, \beta_M^* \} = \argmin_{\{ \beta_1, \ldots, \beta_M \}} \mathbb{E}_{\alpha \sim p_{\Alpha}(\cdot)} \big[\mathcal{L}(\alpha;\beta_1, \ldots, \beta_M) \big],
\end{align}
so that the expected projection error is minimized, where the expectation is taken over the target function parameter $\alpha$.

Determining the optimum $\{ \beta_m^* \}_{m=1}^{M}$ individually can be intractably challenging.
Therefore, we take an alternative approach of treating each $\beta_m$ as a random variable such that an individual pole $\beta_m$ is sampled i.i.d. from the PDF $p_{\Beta}(\cdot)$.
Instead of finding the optimum reservoir weights individually, we attempt to find the optimum probability distribution in terms of its PDF $p_{\Beta}^{*}(\cdot)$, from which the poles $\{ \beta_1, \ldots, \beta_M \}$ can be ``configured'' by sampling from $p_{\Beta}^{*}(\cdot)$ in an i.i.d. manner. 
Therefore, the reservoir optimization problem can be reformulated as determining the optimum PDF $p_{\Beta}^{*}(\cdot)$ of the ESN pole distribution which satisfies
\begin{align}
    &p_{\Beta}^{*}(\cdot) \nonumber \\ 
    &= \argmin_{p_{\Beta}(\cdot)} \mathbb{E}_{ \{\beta_1, \ldots, \beta_M  \} \overset{\text{i.i.d.}}{\sim} p_{\Beta}(\cdot)} \bigg[ \mathbb{E}_{\alpha \sim p_{\Alpha}(\cdot)} \big[\mathcal{L}(\alpha ; \beta_1, \ldots, \beta_M) \big] \bigg].
    \label{eq:original_objective_function}
\end{align}
In the next section, we describe a method of solving this optimization problem by using a local approximation.
\begin{figure}[htbp]
    \centering    \includegraphics[width=\linewidth]{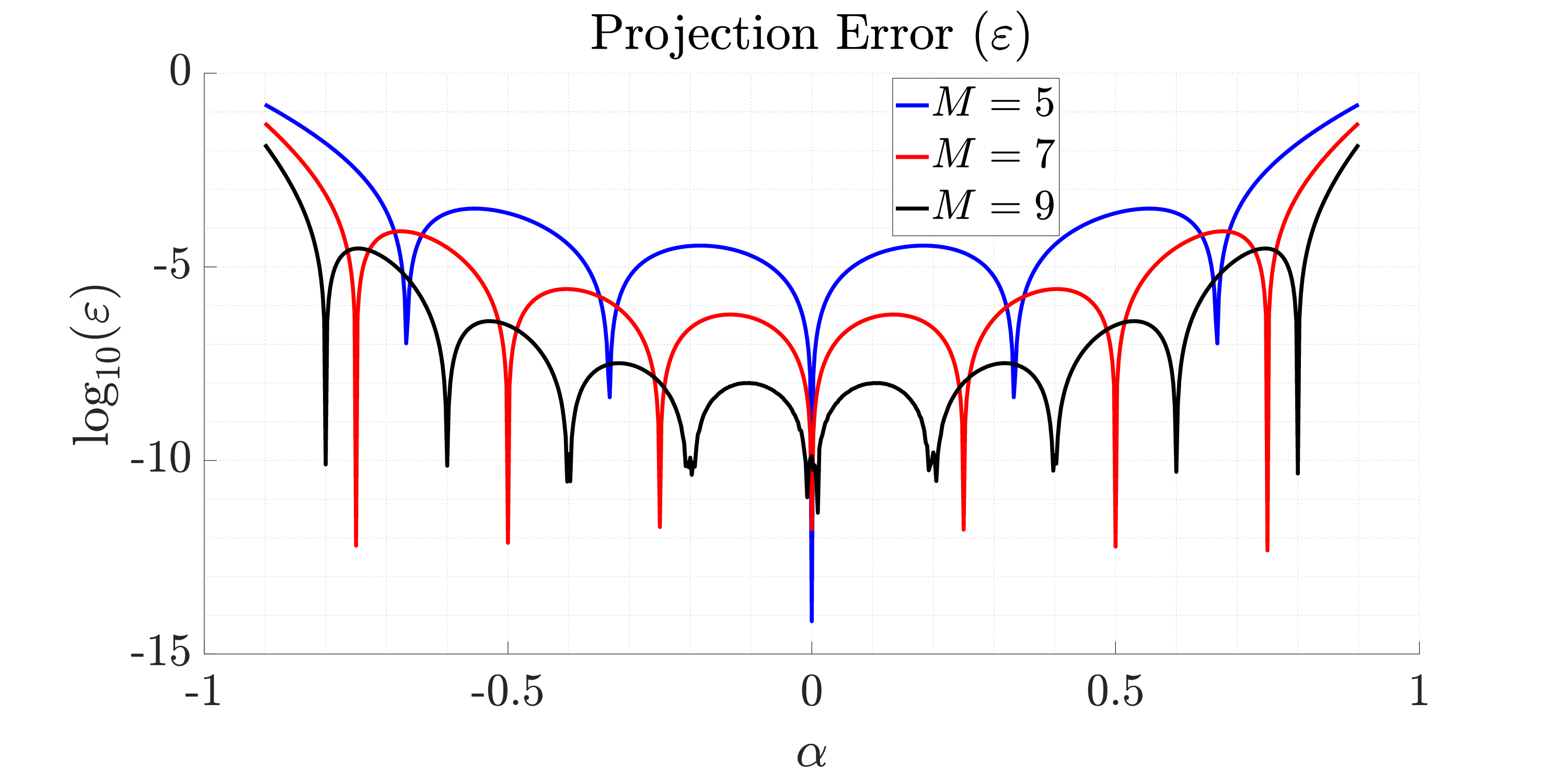}
    \caption{Projection Error ($\varepsilon$) of Eq.~\eqref{eq:simplified_projection_error} versus $\alpha$ for $M=5,7,9$ ESN poles evenly spaced in $(-1,1)$. The local minima represent the locations of the poles $\{\beta_m \}_{m=1}^M$ in each case.}
    \label{fig:Projection_error_M9}
\end{figure}

\section{Reservoir Optimization}
\label{sec:reservoir_optimization}
\subsection{Nearest Neighbors Approximation}
\label{sec:nearest_neighbors_approximation}
As $M \rightarrow \infty$, the projection error $\mathcal{L}(\alpha; \beta_1, \ldots, \beta_M)$ can be estimated by making a ``nearest neighbors approximation''.
This approximation states that in the neighborhood of a given $\alpha$, the approximation error due to $\{\beta_m \}_{m=1}^{M}$ is dominated by the two ESN poles closest to $\alpha$. 
In this treatment, we assume that $\alpha \sim \mathcal{U}(-\alpha_0,\alpha_0) \overset{\mtrian}{=} p_{\Alpha}(\cdot)$, where $0< \alpha_0 < 1$. 
Then, the nearest neighbors approximation states that
\begin{align}
    \mathcal{L}(\alpha; \beta_1, \ldots, \beta_M) \approx \widetilde{\mathcal{L}}(\alpha; \beta_1, \ldots, \beta_M),
    \label{eq:surrogate_loss}
\end{align}
where the ``surrogate loss'' $\widetilde{\mathcal{L}}$ is defined as
\begin{align}
    \widetilde{\mathcal{L}}(\alpha; \beta_1, \ldots, \beta_M) \overset{\mtrian}{=}  \mathcal{L} (\alpha; \beta^{(1)}, \beta^{(2)}).
    \label{eq:neighborhood_loss}
\end{align}
Here, $\beta^{(1)}$ and $\beta^{(2)}$ are the two ESN poles that are closest to $\alpha$, i.e., its two nearest neighbors, with $\beta^{(1)}, \beta^{(2)} \subset \{\beta_1, \ldots, \beta_M \}$. 
In this treatment, we define a local neighborhood $\mathcal{R}$ as the interval containing $\beta^{(1)}$, $\alpha$ and $\beta^{(2)}$, i.e., it is the interval containing the LTI system pole $\alpha$ and only the two nearest ESN poles $\beta_1$ and $\beta_2$.
With the approximation of Eq.~\eqref{eq:surrogate_loss}, the optimization problem can be stated as
\begin{align}
     &p_{\Beta}^{*}(\cdot) \nonumber \\
     &=  \argmin_{p_{\Beta}(\cdot)} \mathbb{E}_{ \{ \beta_m  \} \overset{\text{i.i.d.}}{\sim} p_{\Beta}(\cdot)} \Bigg[ \mathbb{E}_{\alpha \sim p_{\Alpha}(\cdot)} \bigg[\widetilde{\mathcal{L}} \left( \alpha ; \{ \beta_m \}_{m=1}^{M} \right) \bigg] \Bigg].
     \label{eq:optimization_Ltilde}
\end{align}
In the following sequence of steps, we denote $p_{\Beta}^{*}(\cdot)$ as $p_{\Beta}^{*}$ for brevity of notation. If the projection error corresponding to the problem in Eq.~\eqref{eq:optimization_Ltilde} is $\varepsilon_1$, i.e., 
\begin{align}
    &\varepsilon_1 \nonumber \\ &=\min_{p_{\Beta}} \mathbb{E}_{ \{\beta_m  \} \overset{\text{i.i.d.}}{\sim} p_{\Beta}} \bigg[ \mathbb{E}_{\alpha \sim p_{\Alpha}(\cdot)} \big[\widetilde{\mathcal{L}}(\alpha ; \beta_1, \ldots, \beta_M) \big] \bigg], \nonumber \\
    &=\min_{p_{\Beta}} \mathbb{E}_{ \{\beta_m  \} \overset{\text{i.i.d.}}{\sim} p_{\Beta}} \bigg[\sum_{\mathcal{R}} \int_{u \in \mathcal{R}} p_{\Alpha}(u) \widetilde{\mathcal{L}}(\alpha ; \beta_1, \ldots, \beta_M) du  \bigg], \nonumber \\
    &=\min_{p_{\Beta}} \mathbb{E}_{ \{\beta_m  \} \overset{\text{i.i.d.}}{\sim} p_{\Beta}} \bigg[ \sum_{\mathcal{R}} \int_{u \in \mathcal{R}} p_{\Alpha}(u) \mathcal{L} \left(\alpha; \beta^{(1)}, \beta^{(2)} \right) du \bigg], \nonumber \\
    &\leq \min_{p_{\Beta}} \mathbb{E}_{ \{\beta_m  \} \overset{\text{i.i.d.}}{\sim} p_{\Beta}} \bigg[ \sum_{\mathcal{R}} \int_{u \in \mathcal{R}} p_{\Alpha}(u)\cdot \sup_{\alpha \in \mathcal{R}} \mathcal{L} \left(\alpha; \beta^{(1)}, \beta^{(2)} \right) du \bigg],    \nonumber \\
    &=\min_{p_{\Beta}} \mathbb{E}_{ \{\beta_m \} \overset{\text{i.i.d.}}{\sim} p_{\Beta}} \bigg[ \sum_{\mathcal{R}}  \Pr(\alpha \in \mathcal{R}) \cdot \sup_{\alpha \in \mathcal{R}} \mathcal{L} \left(\alpha; \beta^{(1)}, \beta^{(2)} \right) \bigg],
    \label{eq:min_max_optimization_objective}
\end{align}
where the dummy variable $u$ denotes a particular realization of the random variable $\alpha$. Since $\Pr(\alpha \in \mathcal{R})$ is constant regardless of the location of the small neighborhood $\mathcal{R}$ in the entire range of $[-\alpha_0, \alpha_0]$ for $\alpha \sim p_{\Alpha}(\cdot)$, the optimization problem can be stated as the min-max formulation given by
\begin{align}
    p_{\Beta}^{*} = \argmin_{p_{\Beta}} \mathbb{E}_{ \{\beta_m \} \overset{\text{i.i.d.}}{\sim} p_{\Beta}} \bigg[\sum_{\mathcal{R}} \sup_{\alpha \in \mathcal{R}} \big[ \mathcal{L} (\alpha; \beta^{(1)}, \beta^{(2)}) \big] \bigg].
    \label{eq:final_optimization_objective}
\end{align}
As we shall see in the next section, obtaining the exact expression for $\sup_{\alpha \in \mathcal{R}} \big[ \mathcal{L} (\alpha; \beta^{(1)}, \beta^{(2)}) \big]$ can be intractably challenging. Instead, we derive an upper bound $B_{\varepsilon}^{(\mathcal{R})}$ for this term so that $\sup_{\alpha \in \mathcal{R}} \big[ \mathcal{L} (\alpha; \beta^{(1)}, \beta^{(2)}) \big] \leq B_{\varepsilon}^{(\mathcal{R})}$.
Therefore, we define the ``optimum'' distribution $p_{\Beta}^{*}(\cdot)$ as that which solves the following optimization problem:
\begin{align}
    p_{\Beta}^{*}(\cdot) = \argmin_{p_{\Beta}(\cdot)} \mathbb{E}_{ \{\beta_m \} \overset{\text{i.i.d.}}{\sim} p_{\Beta}} \bigg[\sum_{\mathcal{R}} B_{\varepsilon}^{(\mathcal{R})} \bigg].
\end{align}

In the next section, we derive an expression for $B_{\varepsilon}^{(\mathcal{R})}$.

\subsection{Error Bound with Nearest Neighbors Approximation}
\label{sec:nearest_neighbor_optimization}
We analyze the behavior of the error of Eq.~\eqref{eq:simplified_projection_error} in the small neighborhood $\mathcal{R}$ where the two nearest neighbors of $\alpha \in \mathcal{R}$, $\beta^{(1)}$ and $\beta^{(2)}$ are denoted simply as $\beta_1$ and $\beta_2$ for clarity of notation and $\beta_1 < \beta_2$ W.L.O.G.
Thus, for the purpose of this analysis, $\mathcal{R}$ is defined as the interval containing $\beta_1$, $\alpha$ and $\beta_2$.
With this disclaimer, in the following analysis, we denote the projection error of Eq.~\eqref{eq:simplified_projection_error} as $\varepsilon_{(2)}$, where the subscript $(2)$ denotes the fact that we are evaluating a $2$-nearest neighbors error in a small neighborhood $\mathcal{R}$.
$\varepsilon_{(2)} \overset{\mtrian}{=} \mathcal{L}(\alpha; \beta_1, \beta_2)$ can be evaluated by substituting for $\mathbf{r}$ and $\mathbf{\Sigma}$ with $M=2$ in Eq.~\eqref{eq:simplified_projection_error}. After further manipulation, it can be obtained as
\begin{align}
    \varepsilon_{(2)} &= 1 
    -\frac{(1-\alpha^2)(1-\beta_1\beta_2)}{(\beta_1-\beta_2)^2}  \Bigg(\frac{(1-\beta_1^2)(1-\beta_1\beta_2)}{(1-\alpha\beta_1)^2} \nonumber \\ 
    &-2\frac{(1-\beta_1^2)(1-\beta_2^2)}{(1-\alpha\beta_1)(1-\alpha\beta_2)}+\frac{(1-\beta_2^2)(1-\beta_1\beta_2)}{(1-\alpha\beta_2)^2} \Bigg).
    \label{eq:error_expanded}
\end{align}
Since our focus is on the small neighborhood $\mathcal{R}$, we quantify the density of packing of the two ESN poles $\beta_1$ and $\beta_2$ by defining them around a mid-point $\beta_c$ as $\beta_1 = \beta_c - \Delta$ and $\beta_2 = \beta_c + \Delta$, where $\Delta \geq 0$ and $\beta_c \overset{\mtrian}{=} \frac{1}{2}(\beta_1 + \beta_2)$.

We are interested in the trend followed by the maximum value of this error within $\mathcal{R}$ as a function of $\Delta$. 
However, obtaining an expression for the true maximum error $\varepsilon_{(2)}^{(\mathrm{max})}$ by finding the stationary point inside $\mathcal{R}$ can be intractably tedious.
Therefore, instead of finding $\varepsilon_{(2)}^{(\mathrm{max})}$, we attempt to find an upper bound $B_{\varepsilon}$ on $\varepsilon_{(2)}^{(\mathrm{max})}$.
With this final goal, we state the following proposition.
\begin{proposition}
\label{prop1}
    An upper bound on the maximum error in $\mathcal{R}$ is given by 
    \begin{align}
        B_{\varepsilon} = \varepsilon_{(2)}^{\mathrm{(mid)}} + \Delta \left| \frac{\partial \varepsilon_{(2)}}{\partial \alpha}\bigg\rvert_{\alpha = \beta_c} \right|,
        \label{eq:error_bound_expression}
    \end{align}
    where $\varepsilon_{(2)}^{\mathrm{(mid)}}\overset{\mtrian}{=} \varepsilon_{(2)}(\beta_c)$.
\end{proposition}

\begin{figure}[htbp]
    \centering  \includegraphics[width=\linewidth]{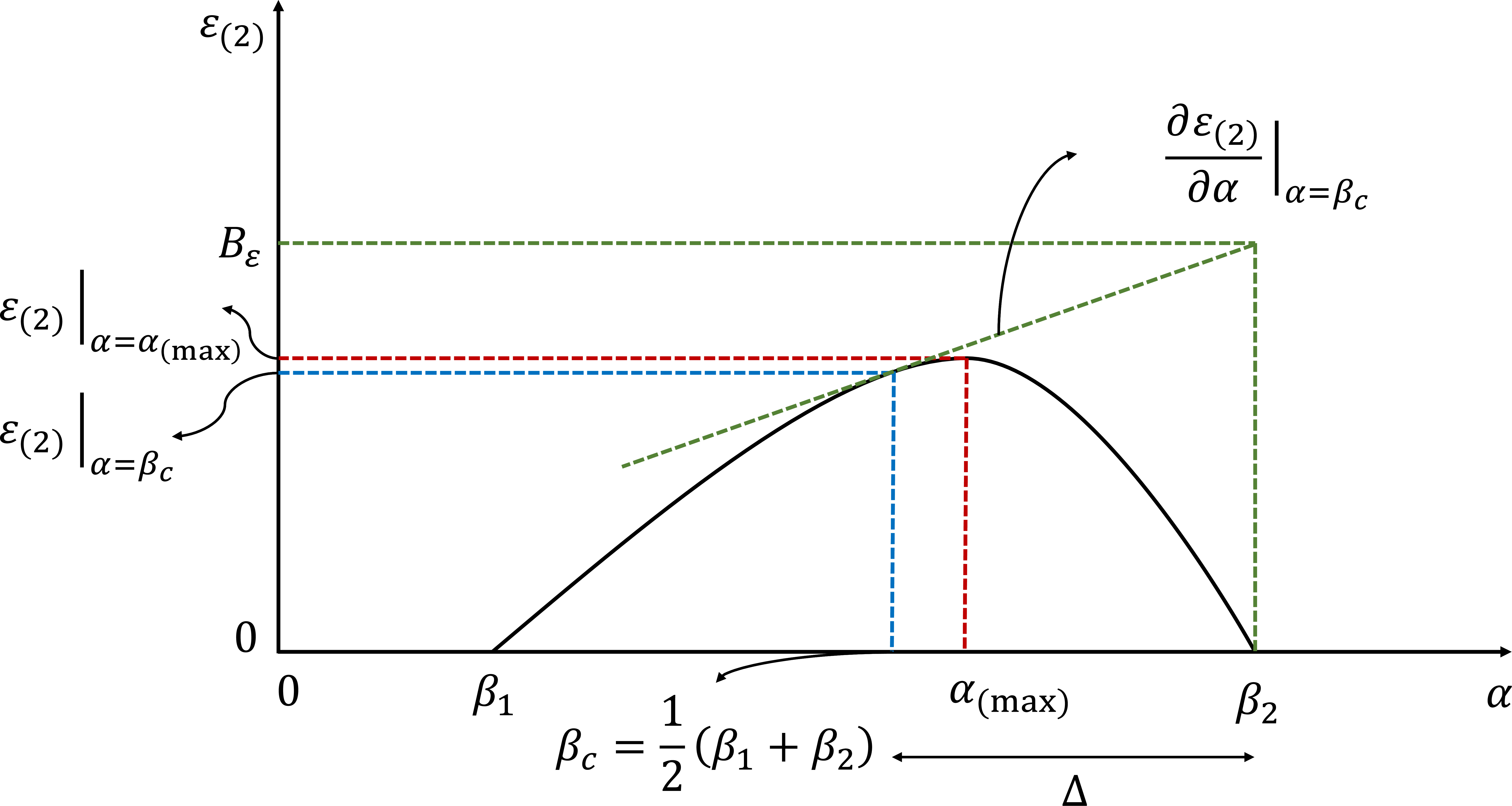}
    \caption{Nearest Neighbors Approximation: Projection error $\varepsilon_{(2)}$ plotted in the neighborhood $\mathcal{R}$ with poles $\beta_1$ and $\beta_2$ on its edges.}
    \label{fig:proposition1}
\end{figure}

This can be seen with the aid of Fig.~\ref{fig:proposition1} which plots the projection error $\varepsilon_{(2)}$ with the nearest neighbors approximation for $\beta_1$, $\beta_2 >0$ W.L.O.G. It can be observed that $B_{\varepsilon}$ of Eq.~\eqref{eq:error_bound_expression} is one of the possible upper bounds on the true maximum error $\varepsilon_{(2)}\rvert_{\alpha = \alpha_{\mathrm{max}}}$. 
Since $\varepsilon_{(2)}$ is a concave function of $\alpha$ and has exactly one local maximum inside $\mathcal{R}$, the claim of Proposition~\ref{prop1} always holds within $\mathcal{R}$ which is bounded by exactly one ESN pole on either side.
Then, $\varepsilon_{(2)}$ evaluated at $\beta_c$ is given in the following lemma.
\begin{lemma}
\label{lemma:error_mid}
    The $2$-nearest neighbors-based projection error $\varepsilon_{(2)}$, evaluated at the mid-point, i.e., $\alpha = \beta_c$ of the small neighborhood $\mathcal{R}$ is given by
    \begin{align}
        \varepsilon_{(2)}^{(\mathrm{mid})} = \frac{1}{(1-\beta_c^2)^4} \Delta^4 + O(\Delta^6).
        \label{eq:error_mid}
    \end{align}
\end{lemma}
The complete proof of this result is provided in Appendix~\ref{appendix:alpha_mid_scaling_law}.
This is an important outcome, indicating that the neighborhood error has a power scaling law with $\Delta$ given by $\Delta^4$. 
Similarly, an expression for $\frac{\partial \varepsilon_{(2)}}{\partial \alpha}\rvert_{\alpha = \beta_c}$ is given in the following lemma.
\begin{lemma}
\label{lemma:error_bound}
    The rate of change of $\varepsilon_{(2)}$ in $\mathcal{R}$, evaluated at the mid-point $\alpha=\beta_c$ is given by
    \begin{align}
        \frac{\partial \varepsilon_{(2)}}{\partial \alpha}\bigg\rvert_{\alpha = \beta_c} = \frac{4\beta_c}{(1-\beta_c^2)^5} \Delta^4 + O(\Delta^6).
        \label{eq:derivative_mid}
    \end{align}
\end{lemma}
The complete derivation for this result is given in Appendix~\ref{appendix:derivative_mid}.
With the results of Lemma~\ref{lemma:error_mid} and Lemma~\ref{lemma:error_bound}, we can use Proposition~\ref{prop1} to state the following theorem.

\begin{thm}
    An upper bound on the worst-case (highest) projection error in $\mathcal{R}$ is given by
    \begin{align}
        B_{\varepsilon} = \frac{1}{(1 - \beta_c^2)^4} \Delta^4 + \frac{4|\beta_c|}{(1-\beta_c^2)^5} \Delta^5 + O(\Delta^7).
        \label{eq:training_loss_scaling_law}
    \end{align}
    \label{thm:training_loss_scaling_law}
\end{thm}
This result follows from directly substituting Eq.~\eqref{eq:error_mid} and Eq.~\eqref{eq:derivative_mid} in Eq.~\eqref{eq:error_bound_expression} of Proposition~\ref{prop1}.
Note that a tighter bound on the true maximum error $\varepsilon_{(2)}^{(\mathrm{max})}$ can be obtained by evaluating the RHS of Eq.~\eqref{eq:error_bound_expression} at an $\alpha=\alpha^*$ that is closer to the true maximizing point $\alpha_{(\mathrm{max})}$, instead of at the mid-point $\alpha = \beta_c$. 
However, it can be shown that such a tighter bound also exhibits a  minimum dependence of $\Delta^4$.
With either upper bound, the conclusion is that the worst-case projection error in $\mathcal{R}$ obeys a scaling law versus $\Delta$ with the minimum exponent $4$ and no lower than that, i.e., the error scales at least as $\Delta^4$, which is a significant result.

\subsection{Deriving the Optimum ESN Pole Distribution}
\label{sec:optimum_pdf_derivation}
Theorem~\ref{thm:training_loss_scaling_law} expresses an upper bound on the approximation error of an LTI system pole $\alpha$ using only two ESN poles $\beta_1$ and $\beta_2$, as a function of the distance between the poles $\Delta = \frac{|\beta_2 - \beta_1|}{2}$. 
Now, we revert to the problem of approximating $\alpha$ using $M$ ESN poles $\{\beta_m \}_{m=1}^M$ that are ``configured'' by sampling them in an i.i.d. manner from the PDF $p_{\Beta}(\cdot)$.
As $M \rightarrow \infty$, we now define a neighborhood $\mathcal{R}$ as an infinitesimally small interval over $(-\alpha_0, \alpha_0)$ in which the PDF $p_{\Beta}(\cdot)$ is constant with value $p_{\Beta}(\mathcal{R})$.
We denote the length of this interval as $|\mathcal{R} |$.
Then, for a particular realization of $\alpha$ say $\alpha^{(\mathcal{R})}$ lying inside $\mathcal{R}$, the nearest neighbors approximation states that the approximation error is given by the two ESN poles say $\beta^{(1, \mathcal{R})}$ and $\beta^{(2, \mathcal{R})}$ that are closest to $\alpha^{(\mathcal{R})}$.
Denote the corresponding minimum distance between them as $\Delta^{(\mathcal{R})} = \frac{|\beta^{(2, \mathcal{R})} - \beta^{(1, \mathcal{R})}  |}{2}$.
The upper bound on the highest error in $\mathcal{R}$ is given by Eq.~\eqref{eq:training_loss_scaling_law}, which we write as $B_{\varepsilon}^{(\mathcal{R})}(\Delta{(\mathcal{R})})$.
Then, the contribution of this particular realization $\alpha^{(\mathcal{R})}$ to the average approximation error across all realizations of $\alpha$ is given by $C^{(\mathcal{R})} = p_{\Alpha}(\alpha^{(\mathcal{R})}) \cdot |\mathcal{R}| \cdot B_{\varepsilon}^{(\mathcal{R})}$.
To satisfy the min-max optimization objective of Eq.~\eqref{eq:min_max_optimization_objective} and thus, that of Eq.~\eqref{eq:final_optimization_objective}, we require that $C^{(\mathcal{R})}$ remain constant across all such neighborhoods, i.e., for any two neighborhoods $\mathcal{R}$ and $\mathcal{R'}$, we require $C^{(\mathcal{R})} = C^{(\mathcal{R'})}$. Since $p_{\Alpha}(\cdot)$ is constant and as $|\mathcal{R}|$ does not depend on $M$ or $\Delta(\mathcal{R})$, we require $B_{\varepsilon}^{(\mathcal{R})} = B_{\varepsilon}^{(\mathcal{R'})}$. Using only the leading terms of Eq.~\eqref{eq:training_loss_scaling_law}, this becomes
\begin{align}
    \frac{\big(\Delta^{( \mathcal R)}\big)^4}{\big(1-\beta_{c}(\mathcal R)^2\big)^4}  &= \frac{\big(\Delta^{(\mathcal R')}\big)^4}{\big(1-\beta_c(\mathcal R')^2\big)^4}, \nonumber \\
    \Rightarrow \big(\Delta^{(\mathcal{R})}\big)^{4} \; &\propto \; (1 - \beta_c(\mathcal{R})^2)^{4}, \nonumber \\
    \Rightarrow \Delta^{(\mathcal{R})}  \; &\propto (1 - \beta_c(\mathcal{R})^2).
\end{align}

Now, $\Delta^{(\mathcal{R})} \; \propto \; \frac{1}{M \cdot p_{\Beta}(\mathcal{R})}$. Thus, the optimum PDF $p_{\Beta}^{*}(\cdot)$ must vary in $\mathcal{R}$ as
\begin{align}
    p_{\Beta}^{*}(\mathcal{R}) \; \propto \; \frac{1}{1 - \beta_{c}(\mathcal{R})^2}.
\end{align}
Since this relationship must hold in every infinitesimally small $\mathcal{R}$, we can write the PDF $p_{\Beta}^{*}(\cdot)$ in terms of the realization $\beta$ of the random variable representing an ESN pole. Hence, we replace $\beta_{c}(\mathcal{R})$ with $\beta$ to write the optimum $p_{\Beta}^{*}(\beta)$ for the ``global'' allocation of ESN poles as
\begin{align}
    p_{\Beta}^{*}(\beta) = \frac{1}{C} \frac{1}{(1 - \beta^2)},
    \label{eq:optimal_beta_distribution}
\end{align}
where the PDF normalizing constant $C$ is found by solving $\int_{-\alpha_0}^{\alpha_0} \frac{C}{1 - \beta^2} d\beta = 1$ for $|\alpha_0| < 1$, giving $C = \log(\frac{1+\alpha_0}{1-\alpha_0})$. 
As an example for $\alpha_0=0.95$,
$C = 3.6636$ and the optimum ESN pole (reservoir weight) distribution is
\begin{align}
    p_{\Beta}^{*}(\beta) = \frac{0.273}{1- \beta^2}.
\end{align}

\begin{figure}[htbp]
    \centering    \includegraphics[width=\linewidth]{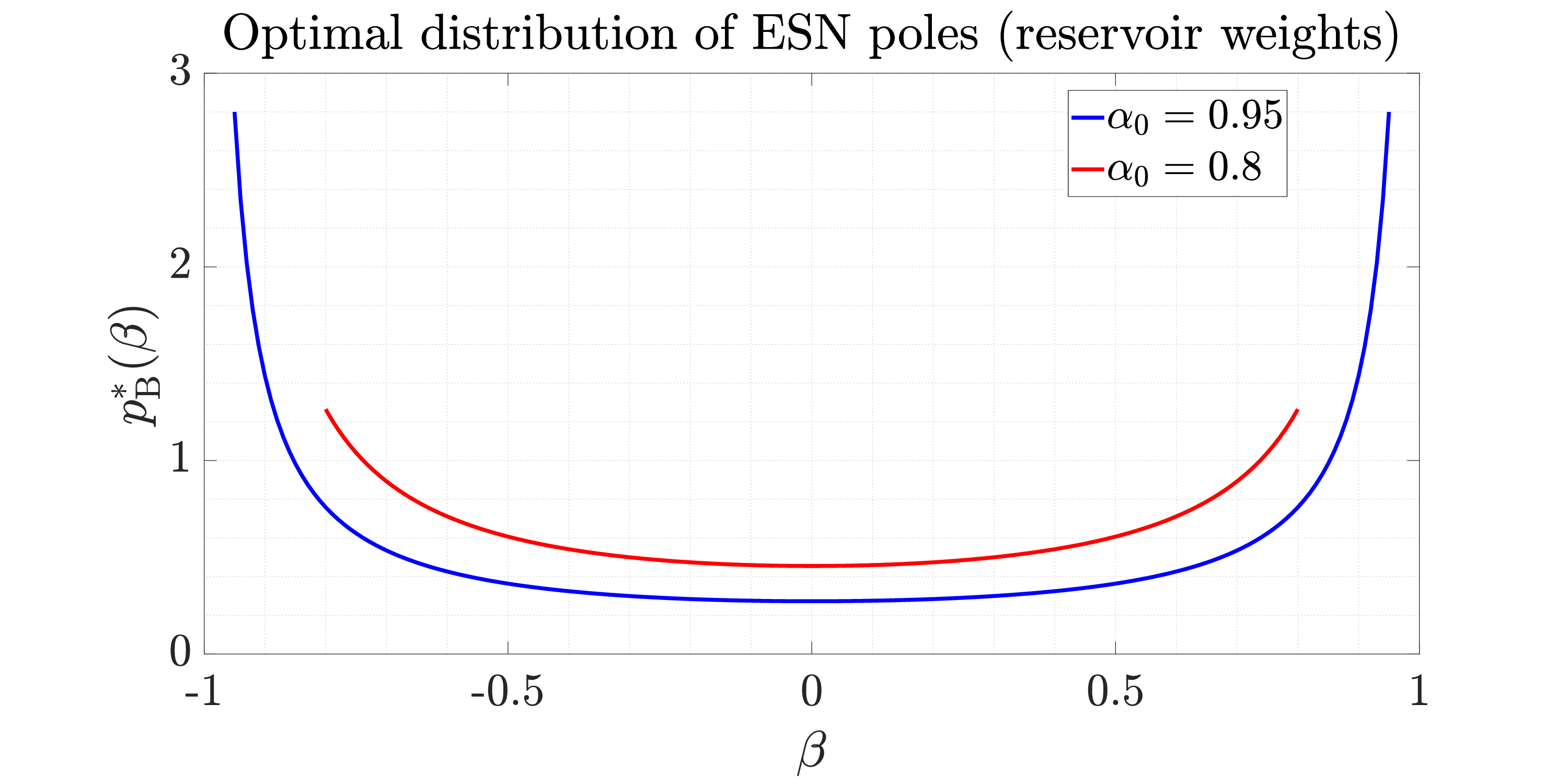}
    \caption{Optimum PDF $p_{\Beta}^{*}(\beta)$ curves for $\alpha_0 = 0.95$ and $\alpha_0 = 0.8$.}
    \label{fig:optimal_distribution}
\end{figure}

The optimum PDF curves for $\alpha_0 = 0.95$ and $\alpha_0 = 0.8$ are plotted in Fig.~\ref{fig:optimal_distribution}.
Recognizing that $\Delta \, \propto \, \frac{1}{M}$, where $M$ is the number of neurons in the reservoir, Theorem~\ref{thm:training_loss_scaling_law} provides a practical scaling law for the ESN projection error and by extension, the training loss, i.e., $\varepsilon_{(2)}^{(\mathrm{max})}\: \propto \: \frac{1}{M^4}$. 
Such a direct scaling relationship of the training loss as a function of the model size is currently not available for more traditional neural network architectures.

\subsection{Incorporating Prior Distributions on pole of LTI system}
In the derivation of $p_{\Beta}^{*}(\beta)$ in Sec.~\ref{sec:optimum_pdf_derivation}, the prior distribution of the pole of the unknown LTI system was assumed to be uniform, i.e., $\alpha \sim \mathcal{U}(-\alpha_0, \alpha_0)$. However, the PDF of the optimum distribution for the poles $\{\beta_m \}$ can be adjusted for any other prior distribution of $\alpha$.
This result is stated in the following corollary.
\begin{corollary}
    Given an optimum probability density function (PDF) $p_{\Beta}^{*}(\cdot)$ of the ESN poles $\{\beta_m \}$ for the unknown LTI system pole $\alpha$ distributed uniformly as $\alpha \sim p_{\Alpha}(\cdot) \overset{\mtrian}{=} \mathcal{U}(-\alpha_0, \alpha_0)$, the optimum ESN pole distribution changes to $q_{\Beta}^{*}(\cdot) \; \propto \; p_{\Beta}^{*}(\cdot) \cdot \big(q_{\Alpha}(\cdot) \big)^{1/4}$, if the prior distribution on $\alpha$ changes from $p_{\Alpha}(\cdot)$ to $q_{\Alpha}(\cdot)$.
    \label{corollary:change_in_prior}
\end{corollary}

We provide a sketch of a proof for this result using the same argument as that in Sec.~\ref{sec:optimum_pdf_derivation}.
For $\alpha \sim q_{\Alpha}(\cdot)$, where $q_{\Alpha}(\cdot)$ is a non-uniform PDF, the contribution of a particular realization $\alpha^{(\mathcal{R})}$ to the average approximation error across all realizations of $\alpha$ is now $C^{(\mathcal{R})} = q_{\Alpha}(\alpha^{(\mathcal{R})}) \cdot |\mathcal{R}| \cdot B_{\varepsilon}^{(\mathcal{R})}$.
Optimizing the min-max objective of Eq.~\eqref{eq:min_max_optimization_objective} requires $C^{(\mathcal{R})} = C^{(\mathcal{R'})}$ for any two neighborhoods $\mathcal{R}$ and $\mathcal{R'}$.
However, since $q_{\Alpha}(\cdot)$ is no longer constant across neighborhoods, this becomes
\begin{align}
    q_{\Alpha}(\alpha^{(\mathcal{R})}) \cdot B_{\varepsilon}^{(\mathcal{R})} = 
    q_{\Alpha}(\alpha^{(\mathcal{R'})}) \cdot B_{\varepsilon}^{(\mathcal{R'})}.
\end{align}
Using only the leading terms in $B_{\varepsilon}^{(\mathcal{R})}$ and $B_{\varepsilon}^{(\mathcal{R'})}$, we get
\begin{align}
    q_{\Alpha}(\alpha^{(\mathcal{R})}) \cdot \frac{\big(\Delta^{( \mathcal R)}\big)^4}{\big(1-\beta_{c}(\mathcal R)^2\big)^4}  &= q_{\Alpha}(\alpha^{(\mathcal{R'})}) \cdot \frac{\big(\Delta^{(\mathcal R')}\big)^4}{\big(1-\beta_c(\mathcal R')^2\big)^4}, \nonumber \\
    \Rightarrow \big(\Delta^{( \mathcal R)}\big)^4 \; &\propto \; \frac{\big(1-\beta_{c}(\mathcal R)^2\big)^4}{q_{\Alpha}(\alpha^{(\mathcal{R})})}.
    \label{eq:change_in_prior_proportionality}
\end{align}
Recognizing that $\Delta^{(\mathcal{R})} \; \propto \; \frac{1}{M \cdot q_{\Beta}(\mathcal{R})}$, we get
\begin{align}
    q_{\Beta}^{*}(\mathcal{R}) \; \propto \; \frac{(q_{\Alpha}(\alpha^{(\mathcal{R})}))^{1/4}}{\big(1-\beta_{c}(\mathcal R)^2\big)}.
\end{align}
Thus, the modified optimum PDF $q_{\Beta}^{*}(\cdot)$ for a global allocation of ESN poles can be written in terms of a general pole realization $\beta$, similar to Sec.~\ref{sec:optimum_pdf_derivation} as
\begin{align}
    q_{\Beta}^{*}(\beta) \, &\propto \, \frac{(q_{\Alpha}(\beta))^{1/4}}{(1 - \beta^2)} \, = p_{\Beta}^{*}(\beta) (q_{\Alpha}(\beta))^{1/4},
    \label{eq:modified_optimal_distribution}
\end{align}
giving the result in Corollary~\ref{corollary:change_in_prior}.

Thus, if the system pole $\alpha$ follows a known non-uniform distribution $q_{\Alpha}(\cdot)$, the ``optimum'' distribution to sample the ESN reservoir weights from is not simply $q_{\Alpha}(\cdot)$ itself, but is rather a function of $q_{\Alpha}(\cdot)$ which is further skewed by the universal optimum PDF $p_{\Beta}^{*}(\cdot)$.
This is an important insight which informs that configuring the ESN reservoir weights according to the same distribution as the LTI system pole is in fact sub-optimal and a better initialization strategy exists.

\subsection{Extension to Higher-order LTI Systems}
\label{sec:extension_to_higher_order_lti}
In the preceding sections, we have considered the atomic problem of approximating a first-order IIR system having a single pole using an ESN consisting of randomly selected reservoir weights (poles) and trained output weights. We now generalize the target function to a higher-order LTI system, in particular, a linear combination of first-order poles~\cite{Oppenheim1996}, i.e., $\mathbf{s}_{u} = \sum_{k=1}^{K} v_k \mathbf{s}_{\alpha_k}$, for some weights $v_k \in \mathbb{R}$, $k=1,\ldots,K$. This is written in the transform domain as
\begin{align}
    S_u(z) = \sum_{k=1}^K \frac{v_k}{1 - \alpha_k z^{-1}}.
    \label{eq:higher_order_LTI_z_domain}
\end{align}
We would like to approximate this higher-order system with an ESN consisting of a random collection of poles $\{\beta_{m,k} \}$ corresponding to each system pole realization $\alpha_k$. 
This approximation can be written as
\begin{align}
    S_u(z) \approx \sum_{k=1}^{K} \left( \sum_{m=1}^{M} \frac{W_{m,k}}{1 - \beta_{m,k}}  \right).
\end{align}

Denoting the projection error incurred in approximating each first-order component $\mathbf{s}_{\alpha_k}$ as $\varepsilon_k$, we recognize that an upper bound on $\mathsf{VAR}(\varepsilon_k)$ has been obtained as $\mathsf{VAR}(\varepsilon_k) \leq  B_{\varepsilon_k}$ in Theorem~\ref{thm:training_loss_scaling_law}.
Then, the variance of the total approximation error $\varepsilon$ across all $K$ poles is given by $\mathsf{VAR}(\varepsilon) = \mathsf{VAR}(\sum_{k=1}^{K} \varepsilon_k)$.
Since the LTI system poles $\{\alpha_k \}$ are not independent in general, an upper bound on $\mathsf{VAR}(\varepsilon)$ can be obtained as
\begin{align}
    \mathsf{VAR}(\varepsilon) \leq K^2 \cdot \max_{k} \big( \mathsf{VAR}(\varepsilon_k) \big) \leq K^2 \cdot \max_{k}\big(B_{\varepsilon_k} \big). 
\end{align}
Therefore, the optimum PDF minimizing the approximation error of a single first-order IIR system also minimizes the same for a linear combination of such poles.

\subsection{Reservoir with Random and Sparse Interconnections}
\label{sec:interconnected_reservoir}
The conventional ESN in state-of-the-art practice uses a reservoir that is sparsely connected with randomly weighted interconnections between the constituent neurons.
In the case of non-interconnected neurons, the reservoir weights matrix is $\mathbf{W}_{\mathrm{res}} = \diag \left(\{ \beta_m \}_{m=1}^{M} \right)$.
However, this is not the case for a sparsely interconnected reservoir.
Performing the eigenvalue decomposition of the general sparse (non-diagonal) $\mathbf{W}_{\text{res}}$,
\begin{align}
    \mathbf{W}_{\text{res}} = \mathbf{Q} \mathbf{\Lambda} \mathbf{Q}^{-1},
    \label{eq:eigendecomposition}
\end{align}
where $\mathbf{Q} \in \mathbb{C}^{M \times M}$ is the matrix containing the eigenvectors of $\mathbf{W}_{\text{res}}$.
For a non-interconnected reservoir, $\mathbf{W}_{\text{res}} = \mathbf{\Lambda}$ and $\mathbf{Q} = \mathbf{I}_M$. 
On the other hand, for a random and sparsely interconnected reservoir, the elements of $\mathbf{W}_{\text{res}}$ induce a corresponding distribution in $\mathbf{\Lambda}$ such that the elements of $\mathbf{\Lambda}$ may no longer be independent~\cite{edelman_rao_2005}.
However, the projection error due to a general random sparsely interconnected reservoir ESN will always be lower bounded by the projection error due to a non-interconnected reservoir with its weights sampled i.i.d. from $p_{\Beta}^{*}(\cdot)$.
Although $p_{\Beta}^{*}(\cdot)$ has been derived for the case of non-interconnected neurons, we will show in this section that even with random and sparse (weighted) interconnections between the neurons, where the recurrent and interconnection weights are drawn from a uniform distribution, the projection error in this case is still lower bounded by the projection error with $\{ \beta_m \} \overset{\text{i.i.d.}}{\sim} p_{\Beta}^{*}(\cdot)$. 
This can be seen by invoking the state update and output equations for the linear ESN, i.e., 
\begin{align}
    \mathbf{x}_{\text{res}}[n] &= \mathbf{W}_{\text{res}} \mathbf{x}_{\text{res}}[n-1] + \mathbf{W}_{\text{in}} \mathbf{x}_{\text{in}}[n] 
    \label{eq:linear_state_update_eqn}
    \\
    \mathbf{x}_{\text{out}}[n] &= \mathbf{W}_{\text{out}} \mathbf{x}_{\text{res}}[n]
    \label{eq:linear_esn_output_eqn}
\end{align}
Substituting Eq.~\eqref{eq:eigendecomposition} in Eq.~\eqref{eq:linear_state_update_eqn}, we get
\begin{align}
    \mathbf{x}_{\text{res}}[n] &= \mathbf{Q} \mathbf{\Lambda} \mathbf{Q}^{-1} \mathbf{x}_{\text{res}}[n-1] + \mathbf{W}_{\text{in}} \mathbf{x}_{\text{in}}[n], \nonumber \\
    \Rightarrow 
    \widetilde{\mathbf{x}}_{\text{res}}[n] &= \mathbf{\Lambda} \widetilde{\mathbf{x}}_{\text{res}}[n-1] + \widetilde{\mathbf{W}}_{\text{in}} \mathbf{x}_{\text{in}}[n],
\end{align}
where $\widetilde{\mathbf{x}}_{\text{res}}[n] \overset{\mtrian}{=} \mathbf{Q}^{-1} \mathbf{x}_{\text{res}}[n]$ and $\widetilde{\mathbf{W}}_{\text{in}} \overset{\mtrian}{=} \mathbf{Q}^{-1} \mathbf{W}_{\text{in}}$.
Using $\mathbf{Q} \mathbf{Q}^{-1} = \mathbf{I}_M$ in Eq.~\eqref{eq:linear_esn_output_eqn}, we get
\begin{align}
    \mathbf{x}_{\text{out}}[n] = \widetilde{\mathbf{W}}_{\text{out}} \widetilde{\mathbf{x}}_{\text{res}}[n],
\end{align}
where $\widetilde{\mathbf{W}}_{\text{out}} = \mathbf{W}_{\text{out}} \mathbf{Q}$.

Thus, a general \emph{linear} ESN with random and sparse interconnections between its reservoir neurons can be diagonalized and the analysis for its optimization is the same as that for a reservoir without interconnections, i.e., for $\mathbf{W}_{\text{res}} = \mathbf{\Lambda}$.
We will empirically show in Sec.~\ref{sec:numerical_evaluation} that a linear reservoir with random interconnections does not provide additional performance gain and is still bounded by the performance of the non-interconnected reservoir ESN with weights sampled from the optimal $p_{\Beta}^{*}(\cdot)$.
This conclusion holds in general for reservoirs with linear activation, i.e., the best performance for a reservoir with linear activation will only be achieved for the case of non-interconnected neurons with $\{\beta_m \}$ configured by sampling i.i.d. from $p_{\Beta}^{*}(\cdot)$
In other words, $p_{\Beta}^{*}(\cdot)$ is the optimum PDF to sample $\{ \beta_m \}$ from only when the neurons are not interconnected. 

Studying the impact of nonlinear activation to derive the optimum PDF, even with non-interconnected neurons can be challenging. Although local approximations of the state update equation around the zero state can be obtained using the Jacobian, which is an approach used in stability analysis~\cite{Bianchi2018}, this is generally analytically tractable only for specific activation functions, e.g., the hyperbolic tangent (tanh) function. 
Alternative approaches may include incorporating the nonlinear activation by modeling the state update equation as a higher-order autoregressive process, up to an order that may admit tractable analysis towards the optimum PDF. 
Finally, operator theoretic methods~\cite{Brunton_Kutz_2019} could be a possible solution for handling the nonlinear activation, however their tractability towards deriving the optimum PDF remains to be studied.
The effect of nonlinear activation on random interconnections between neurons will be addressed in our future work.

\section{Training with Limited Samples}
\label{sec:limited_training}
In the preceding sections, we have considered the orthogonal projection of an LTI system's impulse response on to the subspace spanned by the reservoir of the ESN, and solved the problem of finding the optimum basis for this subspace. 
The optimum output weights $\{W_m^* \}$ for the linear combination of these basis functions are given by Eq.~\eqref{eq:optimum_output_weights}. 
However, this makes use of the knowledge of the particular realization of $\alpha$ or viewed alternatively, requires infinitely many samples to learn $\{W_m^* \}$.
In practice, however, we do not observe or know the true model of the system being simulated, but have access to only a limited number of
labeled input-output data samples.
Under this scenario, the output weights $\mathbf{w} \overset{\mtrian}{=} [W_1 \; W_2\, \; \ldots \; W_M ]^T$ are trained with limited training data using the conventional approach of least squares optimization of the $\ell_2$ regression loss.
For a training sequence consisting of input-output pairs $\{ (x_1, y_1), \ldots, (x_L, y_L) \}$, $\mathbf{w}$ is estimated as
\begin{align}
    \widehat{\mathbf{w}} = \left( \boldsymbol{y}^T \mathbf{X}_{\text{res}}^{\dagger} \right)^{T},
\end{align}
where $\boldsymbol{y} \overset{\mtrian}{=} [y_1\: y_2\,  \ldots \: y_L]^T \in \mathbb{R}^{L \times 1}$ is the ground truth and $\mathbf{X}_{\text{res}} \in \mathbb{R}^{M \times L}$ is the reservoir states matrix containing the state vector $\mathbf{x}_{\mathrm{res}}[n]$ from $n=1$ to $n=L$ in its columns.
When multiple sequences are used for training, the training rule is modified as
\begin{align}
    \widehat{\mathbf{w}} = \left( \widebar{\boldsymbol{y}}^T \widebar{\mathbf{X}}_{\mathrm{res}}^{\dagger} \right)^{T},
    \label{eq:training_rule}
\end{align}
where $\widebar{\boldsymbol{y}} \in \mathbb{R}^{N_p L}$ is the concatenated ground truth across $N_p$ training sequences, and $\widebar{\mathbf{X}}_{\mathrm{res}} \in \mathbb{R}^{M \times N_p L}$ is the concatenated reservoir states matrix.
The availability of only a finite number of labeled training data samples leads to the well-known issue of \emph{model selection}. In the context of ESNs, this translates into selecting an optimum reservoir size $M$ such that the test loss is minimized while avoiding an excessively large reservoir size that may lead to overfitting. 
The Akaike Information Criterion (AIC)~\cite{Akaike1998} is a well-known model selection criterion that penalizes large model sizes. 
The main AIC result can be written as
\begin{align}
    &\argmin_{M} D \left( p_{X,Y}(x,y;\alpha)||p_{X,Y}(x,y;\beta_m, W_m) \right) \nonumber \\ 
    &= \argmin_{M} D \left( \widehat{p}_{X,Y}(x,y;\alpha)||p_{X,Y}(x,y;\beta_m, W_m) \right) + \frac{M}{N_p L},
    \label{eq:AIC_equation}
\end{align}
where $p_{X,Y}(x,y;\alpha)$ denotes the true unknown joint distribution with parameter $\alpha$ from which the input-output sample pairs are generated, i.e., the unknown LTI system. $p_{X,Y}(x,y; \beta_m, W_m)$ denotes the joint distribution generated by the ESN model with parameters $\{\beta_m \}$ and $\{W_m \}$ and $D(p||q)$ denotes the Kullback-Leibler (KL) divergence between two probability distributions with PDFs $p(\cdot)$ and $q(\cdot)$.
Since we cannot observe the true joint distribution $p_{X,Y}(x,y;\alpha)$ in practice and only observe a finite number of input-output samples, we only have access to the empirical joint distribution $\widehat{p}_{X,Y}(x,y;\alpha)$.
Thus, the argument of the LHS of Eq.~\eqref{eq:AIC_equation} is representative of the test loss, while the argument of the first term on the RHS of Eq.~\eqref{eq:AIC_equation} is representative of the training loss computed using a finite number $N_p$ of input-output pair sequences, for which a scaling law as a function of $M$ has been derived in Theorem~\ref{thm:training_loss_scaling_law}.
The second term on the RHS $\frac{M}{N_p L}$ represents the overfitting penalty imposed by the AIC.
Combining this observation with the result of Theorem~\ref{thm:training_loss_scaling_law}, we get 
\begin{align}
    \mathcal{L}_{\text{test}} \; \propto \; \frac{1}{M^4} + \frac{M}{N_p L}
\end{align}
With this relationship, we can derive an order for the optimum reservoir size $M^{*}$ which minimizes the test loss $\mathcal{L}_{\text{test}}$. This is obtained by first setting
\begin{align}
    \frac{d \mathcal{L}_{\text{test}}}{dM} \; \propto \; -\frac{4}{M^5} + \frac{1}{N_p L} = 0.
\end{align}
Solving this, we can obtain an order of magnitude for the optimum reservoir size $M^{*}$ as
\begin{align}
    M^{*} = O \left( (N_p L)^{1/5} \right).
\end{align}

Note that this result does not give the exact reservoir size in terms of number of neurons, but is rather an approximation of the order of the optimum reservoir size needed to minimize the testing loss.
Furthermore, the AIC is one of many model selection criteria, e.g., Bayesian Information Criterion (BIC), Generalized Information Criterion (GIC), among others~\cite{Stoica2004}. 
However, such model selection criteria is beyond the scope of this paper. 
A statistical learning theory-inspired model selection criteria for ESN-based multi-antenna wireless symbol detection is developed in our previous work~\cite{JereTCOM2023}.

\section{Numerical Evaluations}
\label{sec:numerical_evaluation}
In this section, we provide numerical evaluations to validate the theoretical results derived in the preceding sections. 
Specifically, our objective is to experimentally verify the result of Theorem~\ref{thm:training_loss_scaling_law} and validate the optimality of the distribution $p_{\Beta}^{*}(\cdot)$ for the reservoir weights under various scenarios.

\subsection{Sampling from the Optimum Distributions}
For the case of uniformly distributed system pole $\alpha$, we use the  Von Neumann rejection sampling method (accept-reject algorithm)~\cite{robert1999monte} to draw i.i.d. samples from the optimal reservoir weights distribution $p_{\Beta}^{*}(\cdot)$, as well as the modified optimum PDF $q_{\Beta}^{*}(\cdot)$ for non-uniformly distributed $\alpha$, as shall be seen in Sec.~\ref{sec:sim_change_in_prior}. 
Alternatively, its empirical form~\cite{Caffo2002} can avoid computing the PDF scaling constant.

\subsection{Projection Error Scaling Law from Theorem~\ref{thm:training_loss_scaling_law}}
\begin{figure}[htbp]
    \centering    \includegraphics[width=\linewidth]{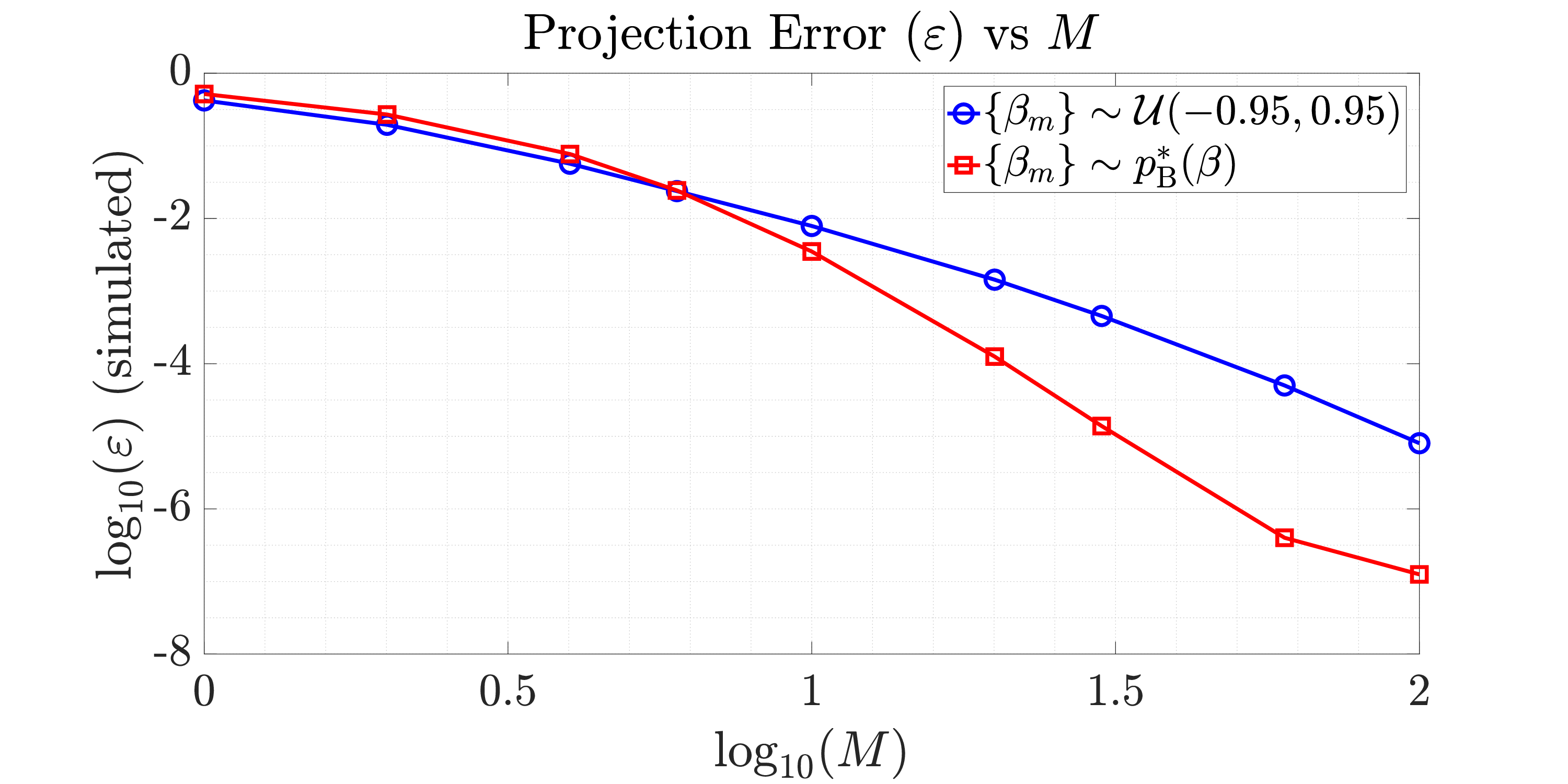}
    \caption{Validation of the scaling law for the projection error ($\varepsilon$) of Eq.~\eqref{eq:simplified_projection_error}}
    \label{fig:scaling_law}
\end{figure}

The main result of Theorem~\ref{thm:training_loss_scaling_law} is a scaling law for the projection error as a function of the reservoir size in neurons. 
This is a key result that also translates to the rate of decrease in the training loss when training a standard \emph{linear} ESN under limited training data.
The projection error $\varepsilon$ of Eq.~\eqref{eq:simplified_projection_error} is simulated over $10^{5}$ Monte-Carlo runs for $\alpha \sim \mathcal{U}(-0.95,0.95)$ for an ESN with a non-interconnected reservoir, i.e., $\mathbf{W}_{\mathrm{res}} = \mathrm{diag}(\{\beta_m \}_{m=1}^M)$.
The resulting plot of the empirical error versus $M$ is shown in Fig.~\ref{fig:scaling_law}.
We can observe that the simulated projection error using reservoir weights $\{\beta_m \}$ configured using the optimum PDF $p_{\Beta}^{*}(\cdot)$ is significantly lower than the error obtained using reservoir weights drawn from $\mathcal{U}(-0.95, 0.95)$. 
Additionally, the empirical $\varepsilon$ approximately displays an $M^{-4}$ dependence when its reservoir weights are configured using $p_{\Beta}^{*}(\beta)$, compared to approximately an $M^{-2}$ dependence displayed when the weights are sampled from $\mathcal{U}(-0.95, 0.95)$, 
indicating a good match between theory and numerical evaluations.

\begin{figure}[htbp]
    \centering    \includegraphics[width=\linewidth]{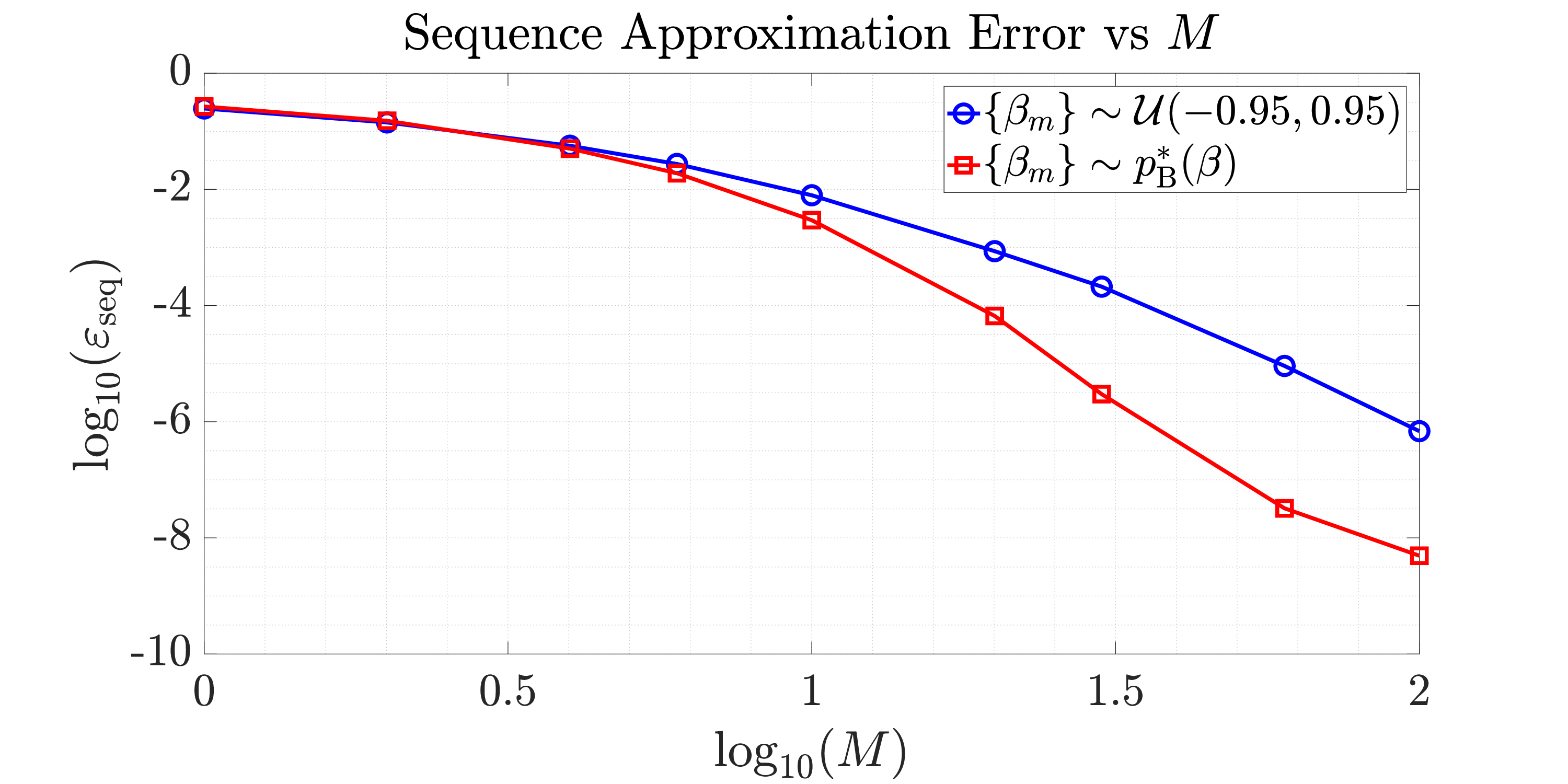}
    \caption{Validation of the scaling law for the sequence approximation error ($\varepsilon_{\mathrm{seq}}$) for sequence length $L=1000$.}
    \label{fig:sequence_approx_error_scaling_law}
\end{figure}

In addition to plotting the projection error, we also validate the scaling law via the empirical sequence approximation error $\varepsilon_{\mathrm{seq}}$, defined as
\begin{align}
    \varepsilon_{\mathrm{seq}} = \frac{1}{N_{\mathrm{sim}} L} \sum_{i=1}^{N_{\mathrm{sim}}} \|\mathbf{y}^{(i)}_{\mathrm{LTI}} - \mathbf{y}^{(i)}_{\mathrm{ESN}} \|_{2}^{2},
\end{align}
where $\mathbf{y}^{(i)}_{\mathrm{LTI}} \in \mathbb{R}^{L}$ and $\mathbf{y}^{(i)}_{\mathrm{ESN}} \in \mathbb{R}^L$ are the sequences each of length $L$ output by the unknown LTI system being simulated and by the ESN approximation respectively in the $i^{\text{th}}$ Monte-Carlo run.
Note that the output weights $\mathbf{w} \in \mathbb{R}^M$ for the sequence approximation task are computed using Eq.~\eqref{eq:optimum_output_weights}, i.e., they are selected as the optimum values $\{ W_m^* \}$ that result from orthogonal projection given the value of the realization of $\alpha$ in each run.
This is plotted in Fig.~\ref{fig:sequence_approx_error_scaling_law} for a sequence length $L=1000$ over $N_{\text{sim}} = 10^5$ runs.
As with the simulated projection error, Fig.~\ref{fig:sequence_approx_error_scaling_law} shows that $\varepsilon_{\mathrm{seq}}$ also exhibits a dependence of approximately $M^{-4}$ for $\{\beta_m \} \sim p_{\Beta}^{*}(\cdot)$ and that of approximately $M^{-2}$ for $\{\beta_m \} \sim \mathcal{U}(-0.95, 0.95)$.
In summary, these numerical evaluations provide strong confirmation for the validity of the derived theoretical optimum distribution of the internal reservoir weights.

\subsection{Training and Test Loss under Limited Training Data}
\label{sec:limited_training_data_evaluation}
\begin{figure}[htbp]
    \centering    \includegraphics[width=\linewidth]{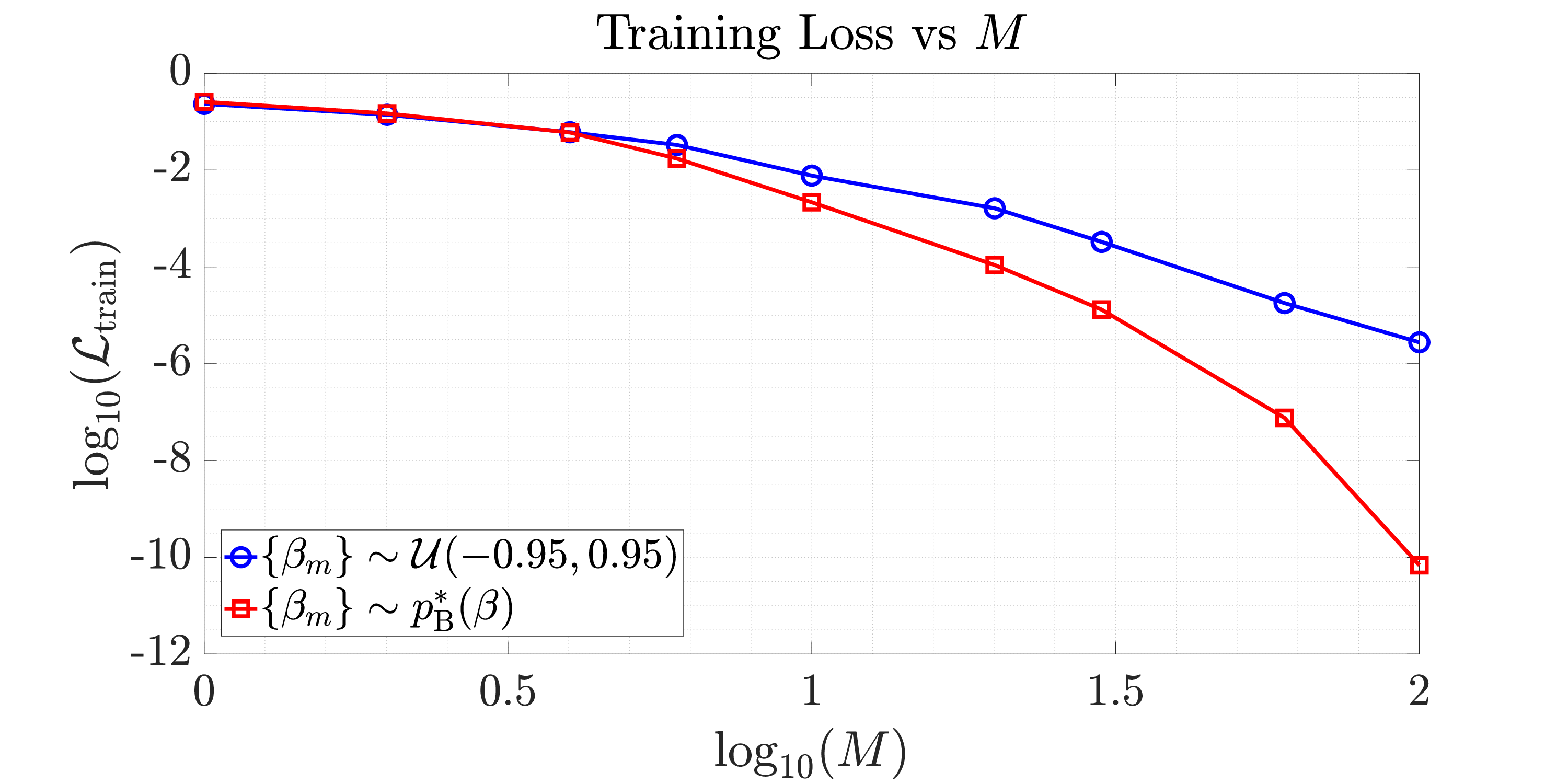}
    \caption{Training loss versus reservoir size $M$ for ESN trained with finite training samples.}    \label{fig:training_loss_vs_M}
\end{figure}

Recall that computing the projection error $\varepsilon$ via Eq.~\eqref{eq:simplified_projection_error} required knowledge of the particular realization of $\alpha$ in each run, or alternatively the availability of infinitely many training samples.
However, with limited training data as in practice, we can verify that a similar scaling trend versus $M$ and improvement in performance in terms of the training and test losses is obtained when configuring the weights using $p_{\Beta}^{*}(\cdot)$ compared to randomly generating them from $\mathcal{U}(-\alpha_0, \alpha_0)$.
To validate this, the linear ESN is trained with $N_p = 1$ training sequence of length $L = 500$ samples, i.e., $\widehat{\mathbf{w}}$ is computed using Eq.~\eqref{eq:training_rule}.
Next, it is tested with $N_d = 10$ test sequences of the same length.
The empirical training loss $\mathcal{L}_{\mathrm{train}} \overset{\mtrian}{=} \frac{1}{N_{\mathrm{sim}} N_p L} \sum_{i=1}^{N_{\mathrm{sim}}} \| \widebar{\mathbf{y}}^{(i)}_{\mathrm{LTI,train}} - \widebar{\mathbf{y}}^{(i)}_{\mathrm{ESN,train}} \|_{2}^{2}$ is plotted in Fig.~\ref{fig:training_loss_vs_M}, where $\widebar{\mathbf{y}}^{(i)}_{\mathrm{LTI,train}} \in \mathbb{R}^{N_p L}$ is the concatenated training output from the LTI system and $\widebar{\mathbf{y}}^{(i)}_{\mathrm{ESN,train}} \in \mathbb{R}^{N_p L}$ is the concatenated ESN output during training respectively in the $i^{\mathrm{th}}$ Monte-Carlo run, with $N_{\mathrm{sim}} = 5\times 10^4$.

\begin{figure}[htbp]
    \centering    \includegraphics[width=\linewidth]{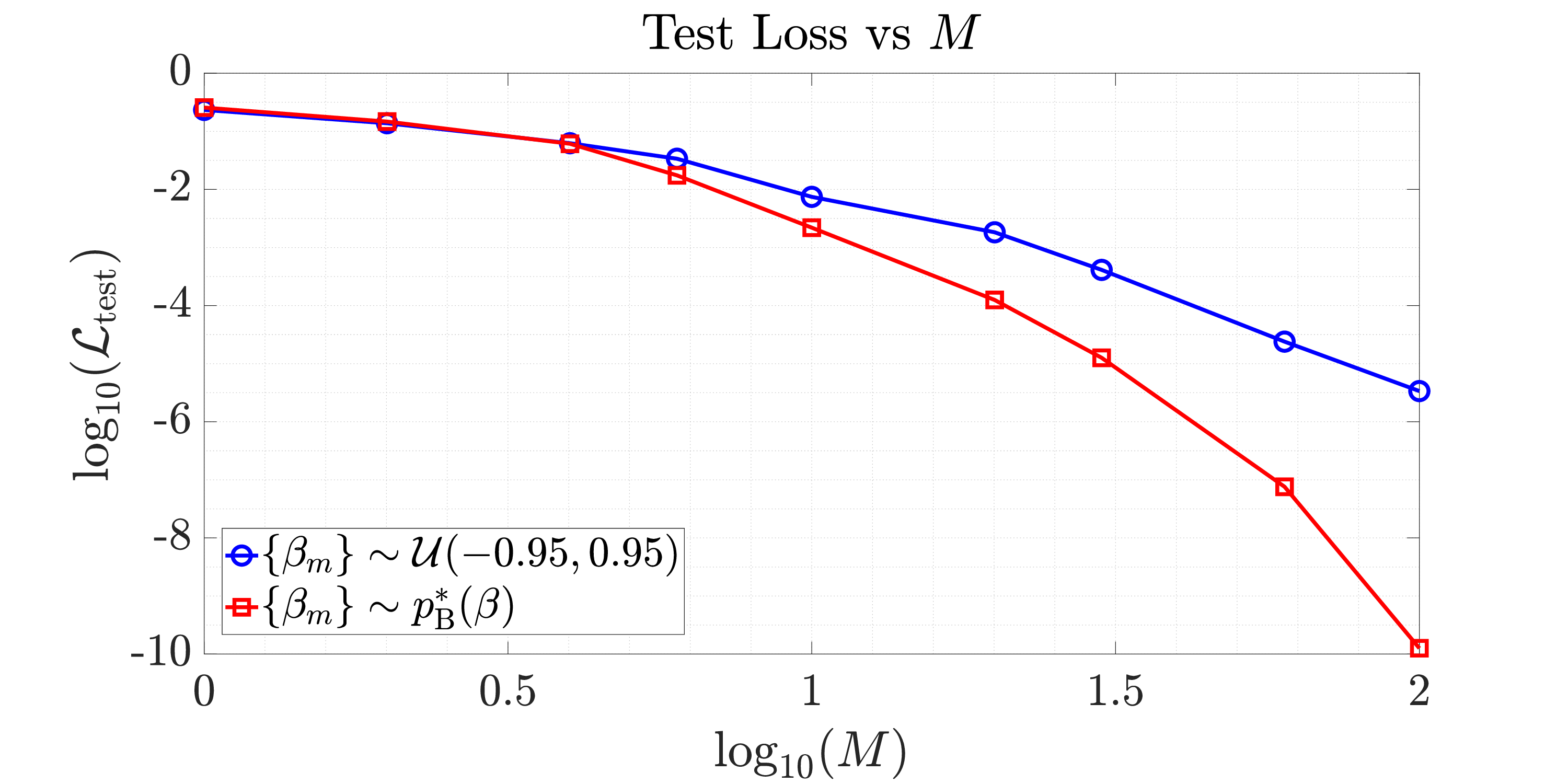}
    \caption{Test loss versus reservoir size $M$ for ESN trained with finite training samples.}    \label{fig:test_loss_vs_M}
\end{figure}

We can observe that the ESN with optimally sampled reservoir weights shows a significantly lower training loss and approximately obeys the $M^{-4}$ scaling law.
The more important practical performance metric, namely the empirical test loss $\mathcal{L}_{\mathrm{test}} \overset{\mtrian}{=} \frac{1}{N_{\mathrm{sim}} N_{d} L} \sum_{i=1}^{N_{\mathrm{sim}}} \| \widebar{\mathbf{y}}^{(i)}_{\mathrm{LTI,test}} - \widebar{\mathbf{y}}^{(i)}_{\mathrm{ESN,test}} \|_{2}^{2}$ is plotted in Fig.~\ref{fig:test_loss_vs_M}, where $\widebar{\mathbf{y}}^{(i)}_{\mathrm{LTI,test}} \in \mathbb{R}^{N_d L}$ is the concatenated LTI system output during test and $\widebar{\mathbf{y}}^{(i)}_{\mathrm{ESN,test}} \in \mathbb{R}^{N_d L}$ is the concatenated ESN output during test respectively in the $i^{\mathrm{th}}$ Monte-Carlo run. 
Therefore, the derived optimum PDF for the reservoir weights can provide up to $4$ orders of magnitude improvement in the test loss at higher reservoir sizes, indicating a huge performance gain that can be achieved without any additional training complexity.
Note that in the simulation of a simple system such as a first-order IIR system, it would typically take a model of a significantly larger size, i.e., reservoir with many more neurons to start observing the overfitting effect in the test loss $\mathcal{L}_{\mathrm{test}}$.

\subsection{Interconnected Linear Activation Reservoir}
In order to validate our finding from Sec.~\ref{sec:interconnected_reservoir} that interconnections between neurons in the reservoir is equivalent to a non-interconnected reservoir with modified input and output weights matrices, we replicate the evaluations of Sec.~\ref{sec:limited_training_data_evaluation}, but with a non-diagonal $\mathbf{W}_{\mathrm{res}}$, i.e., with random and sparse interconnections between the reservoir neurons.
The sparsity of connections is controlled via the hyperparameter `sparsity' (denoted as $\kappa$) which represents the probability of each element of $\mathbf{W}_{\mathrm{res}}$ being $0$.
Furthermore, the spectral radius of $\mathbf{W}_{\mathrm{res}}$ is set to $0.95$, i.e., $\rho(\mathbf{W}_{\mathrm{res}}) = \max | \lambda(\mathbf{W}_{\mathrm{res}})| = 0.95$ for the cases of random and sparsely interconnected reservoirs, with corresponding weights drawn i.i.d. from $\mathcal{U}(-0.95, 0.95)$.
\begin{figure}[htbp]
    \centering    \includegraphics[width=\linewidth]{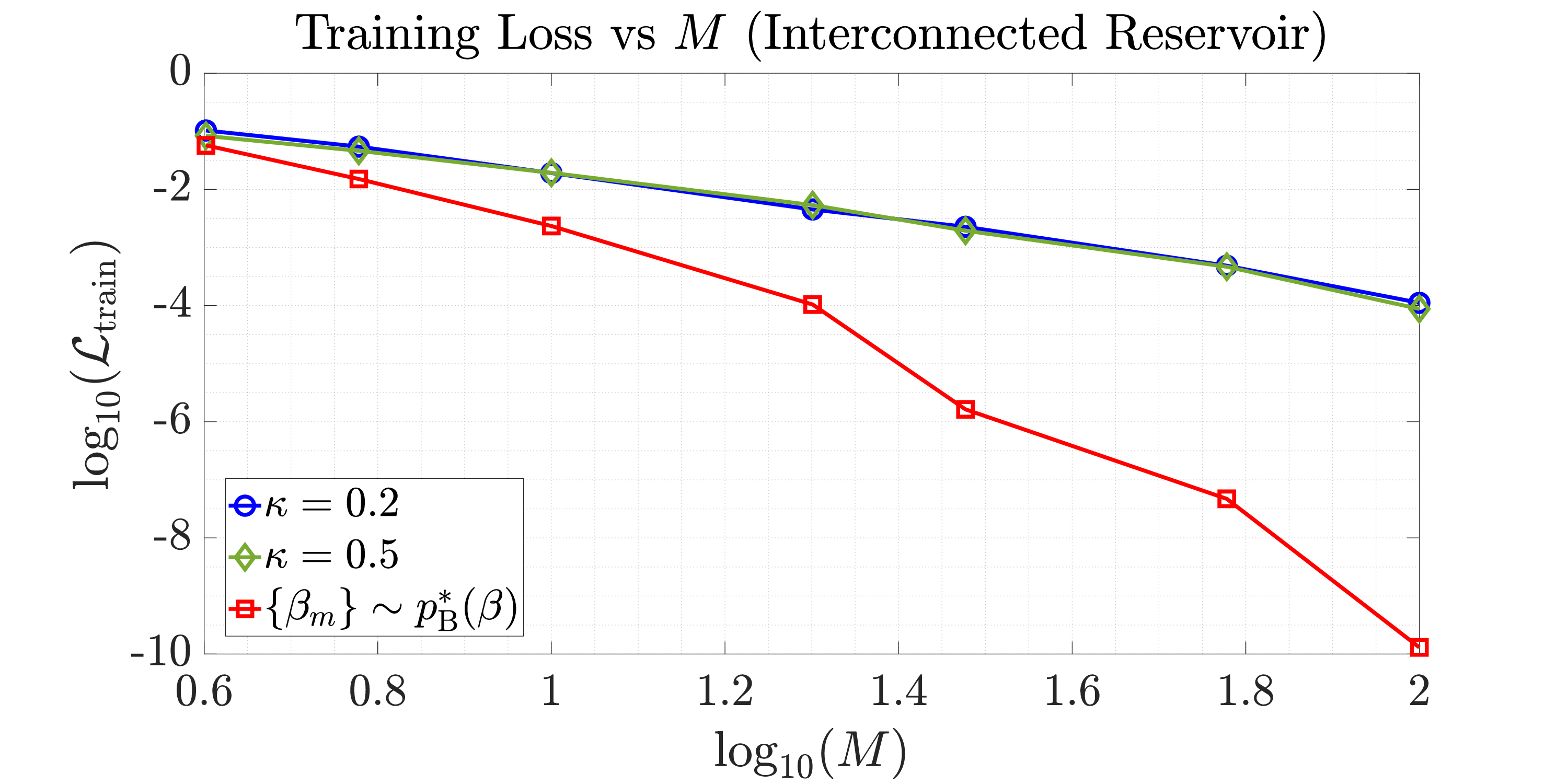}
    \caption{Training loss versus reservoir size $M$ under finite training samples for ESN with random interconnections between neurons.}    \label{fig:interconnected_reservoir_train_loss_vs_M}
\end{figure}

\begin{figure}[htbp]
    \centering    \includegraphics[width=\linewidth]{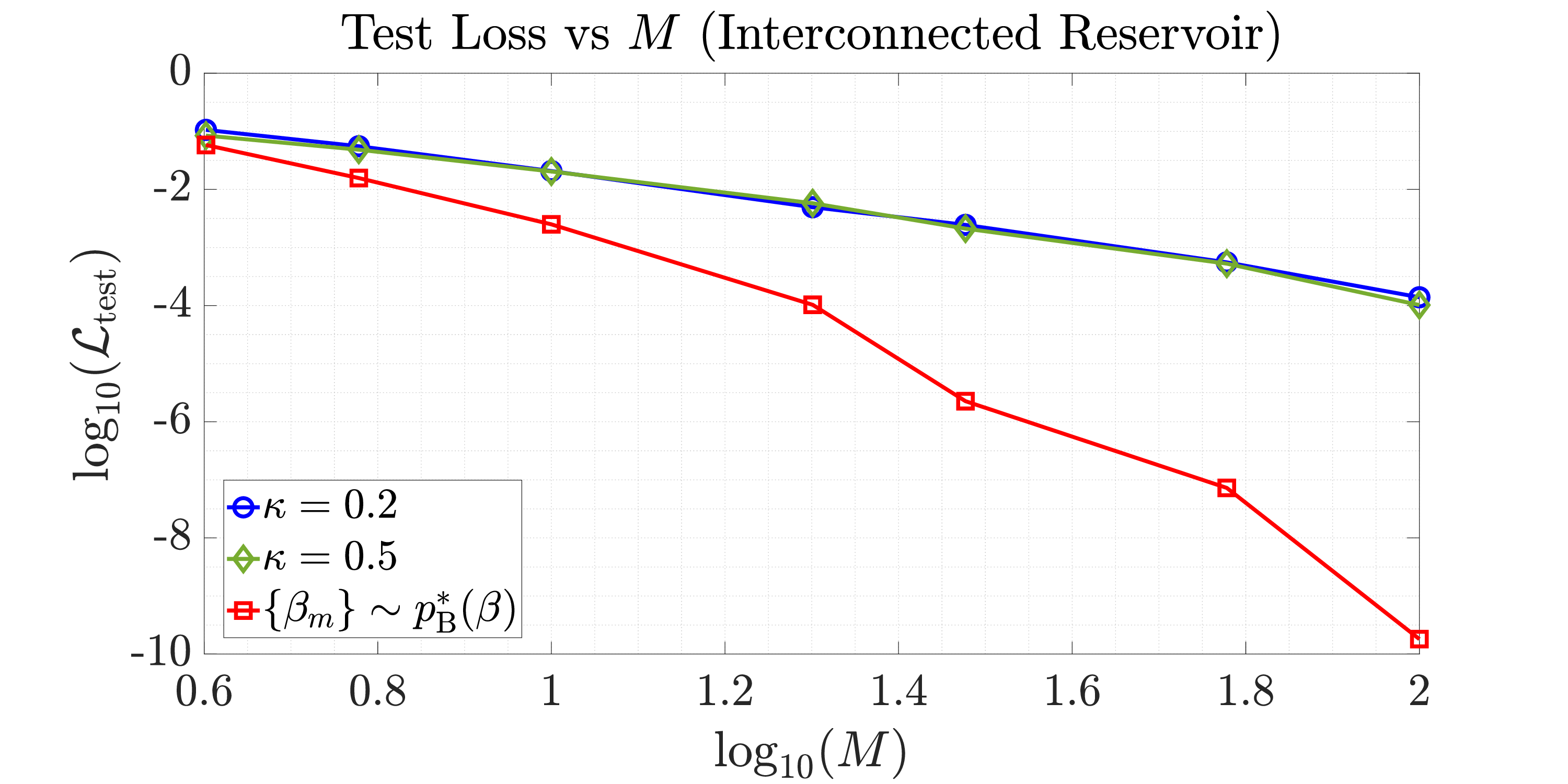}
    \caption{Test loss versus reservoir size $M$ under finite training samples for ESN with random interconnections between neurons.}    \label{fig:interconnected_reservoir_test_loss_vs_M}
\end{figure}
From both Fig.~\ref{fig:interconnected_reservoir_train_loss_vs_M} and Fig.~\ref{fig:interconnected_reservoir_test_loss_vs_M}, we can observe that the ESN with a non-interconnected reservoir with weights configured using the optimum PDF $p_{\Beta}^{*}(\cdot)$ greatly outperforms the ESN model with a sparsely interconnected reservoir with weights randomly generated from $\mathcal{U}(-0.95, 0.95)$, i.e., the state-of-the-art practice.
At higher reservoir sizes, e.g., $M = 100$, we can see a gain of up to $6$ orders of magnitude in the test loss. 
Additionally, for a fixed spectral radius, a change in the sparsity of the reservoir from $\kappa=0.2$ to $\kappa=0.5$ does not result in any observable change in the trends of the training and the test losses.
This confirm our hypothesis from Sec.~\ref{sec:interconnected_reservoir} that for linear activation, random interconnections between neurons do not provide additional performance gain.

\subsection{Simulating change in Prior Distribution of System Pole}
\label{sec:sim_change_in_prior}
\begin{figure}[htbp]
    \centering    \includegraphics[width=\linewidth]{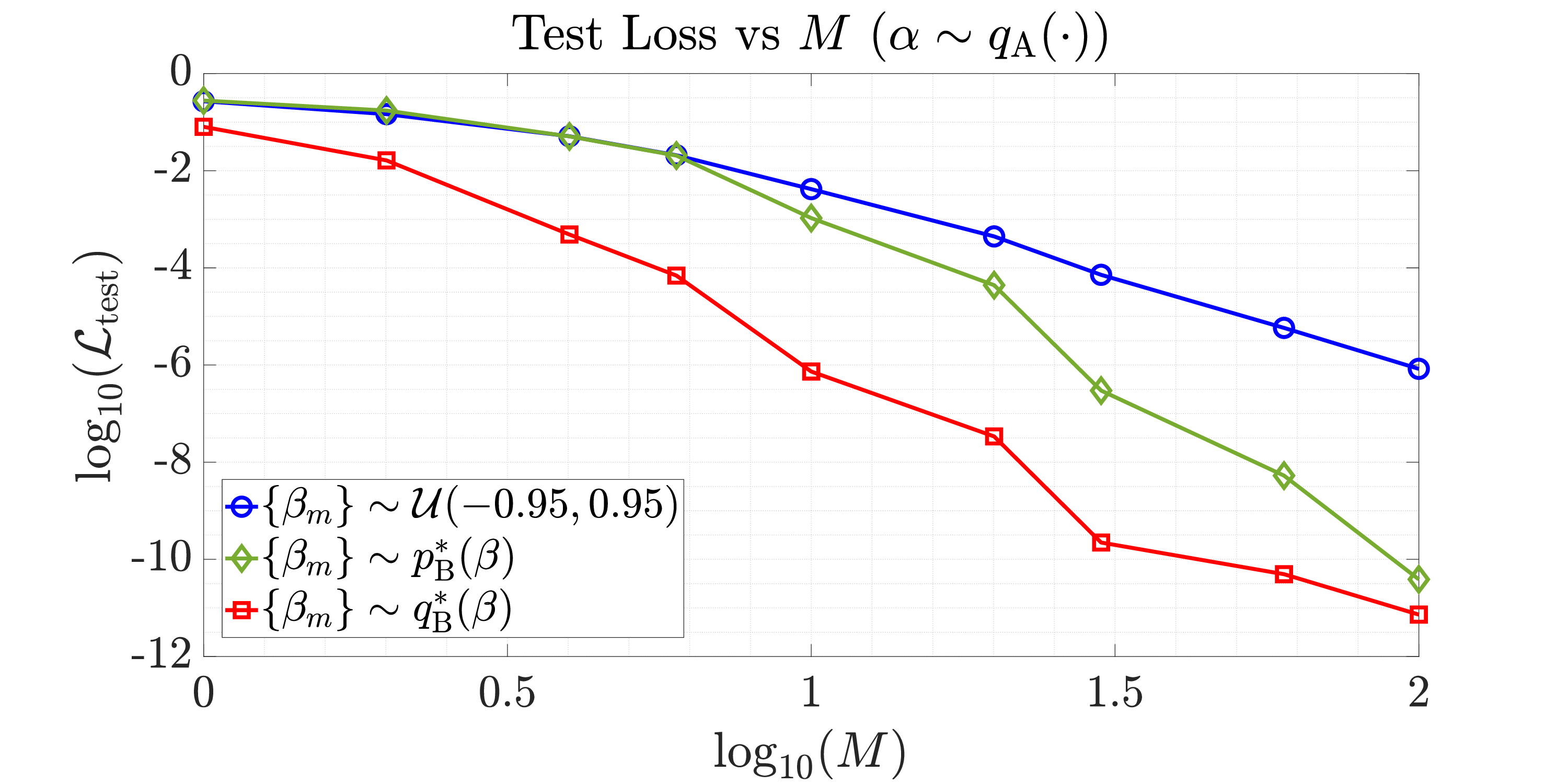}
    \caption{Test loss versus reservoir size $M$ with a changed prior PDF on $\alpha$ given by $q_{\Alpha}(\cdot)$ (non-uniform distribution).}    \label{fig:changed_prior_test_loss_vs_M}
\end{figure}

In this section, the result of Corollary~\ref{corollary:change_in_prior} is validated through simulations. 
Specifically, we consider a changed prior distribution of $\alpha$ given by $q_{\Alpha}(\cdot) \overset{\mtrian}{=} \mathcal{N}(0.7, 10^{-2})$. 
Following the same settings as in the previous sections for data-driven training of the output weights, the ESN reservoir weights are now configured by sampling from the modified optimum PDF $q_{\Beta}^{*}(\cdot)$ using the result of Corollary~\ref{corollary:change_in_prior}.
The corresponding test loss for this experiment is plotted in Fig.~\ref{fig:changed_prior_test_loss_vs_M}.
Furthermore, we also plot the test loss for the case of $\{\beta_m \}$ initialized from $p_{\Beta}^{*}(\cdot)$, which is optimized for $\alpha \sim \mathcal{U}(-0.95, 0.95)$ and not for $\alpha \sim q_{\Alpha}(\cdot)$.
Compared to randomly generating the reservoir weights $\{\beta_m \}$ from $\mathcal{U}(-0.95, 0.95)$, configuring them using $p_{\Beta}^{*}(\cdot)$ or $q_{\Beta}^{*}(\cdot)$ both result in much improved performance.
However, the performance achieved with $q_{\Beta}^{*}(\cdot)$ which is optimized for the modified prior PDF $q_{\Alpha}(\cdot)$ is even better than that using $p_{\Beta}^{*}(\cdot)$ which is optimized for a uniform prior PDF $p_{\Alpha}(\cdot)$.
This clearly validates Corollary~\ref{corollary:change_in_prior} and demonstrates the value in adapting the reservoir initialization strategy to the available domain knowledge.

\subsection{Simulating Higher-order LTI Systems}
\begin{figure}[htbp]
    \centering    \includegraphics[width=\linewidth]{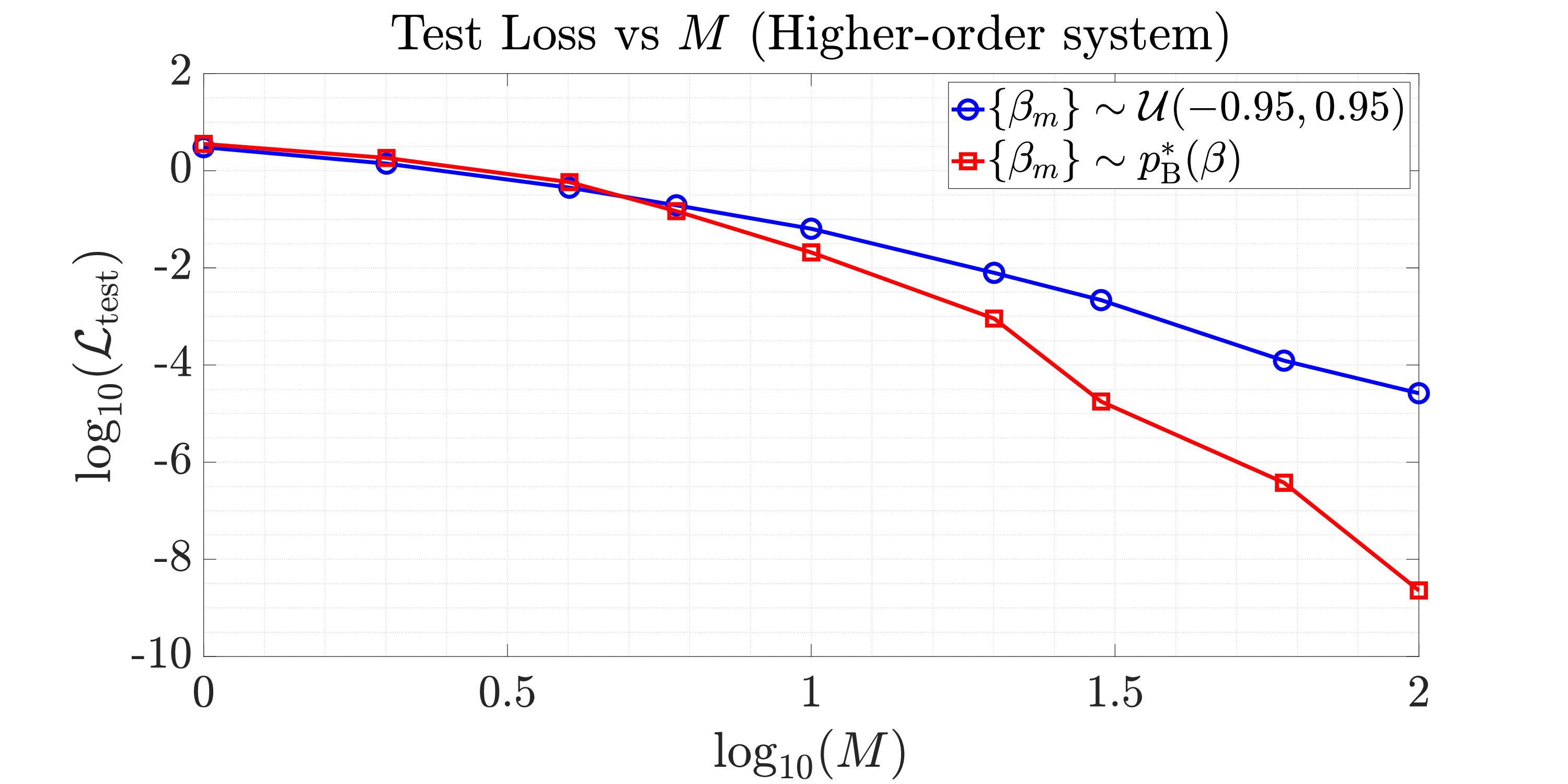}
    \caption{Test loss versus reservoir size $M$ for a $5$-th order LTI system.}    \label{fig:5th_order_lti_sim}
\end{figure}

In this section, we empirically verify the optimality of $p_{\Beta}^{*}(\cdot)$ when approximating higher-order LTI systems of the form given in Eq.~\eqref{eq:higher_order_LTI_z_domain}, i.e., a linear combination of first-order poles. 
Specifically, we consider a $5$-th order system by substituting $K=5$ in $\mathbf{s}_{u} = \sum_{k=1}^{K} {\mathbf{s}}_{\alpha_k}$, where $\{\alpha_k \}$ are sampled i.i.d. from $\mathcal{U}(-0.95, 0.95)$.
The corresponding test loss is plotted in Fig.~\ref{fig:5th_order_lti_sim}.
Similarly to the first-order system approximation task, we can see an improvement of up to $4$ orders of magnitude at moderate to higher reservoir sizes.
This validates the applicability of the derived optimum PDF to higher-order LTI systems.

\section{Conclusion and Future Work}
\label{sec:conclusion}
In this work, we have introduced a clear signal processing approach to understand the echo state network (ESN), a powerful architecture of the Reservoir Computing (RC) family, belonging to the broader class of randomized recurrent neural networks. 
Employing the linear ESN to approximate a simple linear time-invariant (LTI) system, we provide a precise scaling law obeyed by the approximation error and a complete analytical characterization of the optimum probability density function (PDF) that can be used to configure the ESN's reservoir weights, which are otherwise randomly generated in a pre-determined and arbitrary fashion in state-of-the-art practice.
Numerical evaluations demonstrate the optimality of the derived optimum PDF by showing a gain of up to $4$ orders of magnitude at moderate to high reservoir sizes.
This demonstrates the practical applicability and realizable performance gains by virtue of the analysis in this work. 
Extension of this analysis to complex-valued ESNs and developing an understanding of the impact of nonlinear activation is part of future investigation. Additionally, deriving the optimum weights distribution for the wireless channel equalization task given statistical knowledge of the channel is also included in future work.

\begin{appendices}
\section{Proof of Lemma~\ref{lemma:error_mid}}
\label{appendix:alpha_mid_scaling_law}
\begin{proof}
Substituting $\alpha = \beta_c \overset{\mtrian}{=} \frac{1}{2} (\beta_1 + \beta_2)$ in Eq.~\eqref{eq:error_expanded} and the substitution $\beta_1 = \beta_c - \Delta$ and $\beta_2 = \beta_c + \Delta$, we can arrive at the following expression after some manipulation,
\begin{align}
    \varepsilon_{(2)}^{(\mathrm{mid})} &= \frac{\Delta^4}{\big(1 + \beta_c^4 - \beta_c^2 (2 + \Delta^2) \big)^2}, \nonumber \\
    &= \frac{\Delta^4}{(1-\beta_c^2)^4 \left(1 - \frac{2\beta_c^2\Delta^2}{(1-\beta_c^2)^2} + \frac{\beta_c^4\Delta^4}{(1-\beta_c^2)^4} \right) }.
    \label{eq:error_mid_alpha}
\end{align}
To perform a Taylor series expansion up to the second power for the term $C_4^{(\mathrm{mid})}  \overset{\mtrian}{=} \frac{1}{\left(1 - \frac{2\beta_c^2\Delta^2}{(1-\beta_c^2)^2} + \frac{\beta_c^4\Delta^4}{(1-\beta_c^2)^4} \right)}$,
recall the Taylor series expansion for $\frac{1}{1+x}$ for  $x \ll 1$ given by 
\begin{align}
    \frac{1}{1+x} \approx 1 - x + x^2 + O(x^3).
\end{align}
Applying this to $C_4^{(\mathrm{mid})}$, we obtain
\begin{align}
    C_4^{(\mathrm{mid})} &\approx 1 - \left( \frac{\beta_c^4\Delta^4}{(1-\beta_c^2)^4} - \frac{2\beta_c^2\Delta^2}{(1-\beta_c^2)^2} \right) \nonumber \\
    &+ \left(\frac{\beta_c^4\Delta^4}{(1-\beta_c^2)^4} - \frac{2\beta_c^2\Delta^2}{(1-\beta_c^2)^2} \right)^2, \nonumber \\
    &= 1 + \frac{2\beta_c^2\Delta^2}{(1-\beta_c^2)^2} + \frac{3\beta_c^4\Delta^4}{(1-\beta_c^2)^4} + O(\Delta^6).
\end{align}
Using this approximation in Eq.~\eqref{eq:error_mid_alpha}, we arrive at Lemma~\ref{lemma:error_mid},
\begin{align}
    \varepsilon_{(2)}^{(\mathrm{mid})} = \frac{1}{(1-\beta_c^2)^4}\Delta^4 + O(\Delta^6).
\end{align}
\end{proof}

\section{Proof of Lemma~\ref{lemma:error_bound}}
\label{appendix:derivative_mid}
\begin{proof}
With the substitutions $\beta_1 = \beta_c - \Delta$, $\beta_2 = \beta_c + \Delta$ and a sequence of algebraic manipulations, we can arrive at the following expression for the derivative of the neighborhood error w.r.t. $\alpha$, evaluated at the mid-point $\alpha = \beta_c \overset{\mtrian}{=} \frac{1}{2}(\beta_1 + \beta_2)$,
\begin{align}
    \frac{\partial \varepsilon_{(2)}}{\partial \alpha} \bigg\rvert_{\alpha = \beta_c} 
    =  \frac{4\beta_c (1 - \beta_c^2 + \Delta^2)}{\big( (1-\beta_c^2)^2 - \beta_c^2\Delta^2 \big)^3} \Delta^4.
    \label{eq:derivative_expression}
\end{align}
Since a power series expansion for $ \frac{\partial \varepsilon_{(2)}}{\partial \alpha} \big\rvert_{\alpha = \beta_c}$ in terms of $\Delta$ is required, we first obtain a Taylor series expansion for the term $\frac{1}{\left( (1-\beta_c^2)^2 - \beta_c^2\Delta^2 \right)^3}$ as follows. Rewriting this term as
\begin{align}
    &\frac{1}{\big( (1-\beta_c^2)^2 - \beta_c^2\Delta^2 \big)^3} \nonumber \\
    &= \frac{1}{(1-\beta_c^2)^6 \bigg(1 - \frac{\beta_c^6\Delta^6}{(1-\beta_c^2)^6} +\frac{3\beta_c^4\Delta^4}{(1-\beta_c^2)^4} - \frac{3\beta_c^2\Delta^2}{(1-\beta_c^2)^2} \bigg)}.
\end{align}
Next, we perform an expansion for the term 
\begin{align}
    C_{6}^{(\mathrm{bound})} \overset{\mtrian}{=} \frac{1}{1 + \bigg(\frac{3\beta_c^4\Delta^4}{(1-\beta_c^2)^4} - \frac{3\beta_c^2\Delta^2}{(1-\beta_c^2)^2} - \frac{\beta_c^6\Delta^6}{(1-\beta_c^2)^6} \bigg)},
\end{align}
using the Taylor series $\frac{1}{1+x} \approx 1 - x$, for small $x$. Thus, 
\begin{align}
     C_{6}^{(\mathrm{bound})} \approx 1 + \frac{3\beta_c^2\Delta^2}{(1-\beta_c^2)^2} - \frac{3\beta_c^4\Delta^4}{(1-\beta_c^2)^4} + \frac{\beta_c^6\Delta^6}{(1-\beta_c^2)^6}.
     \label{eq:Taylor_series_C6bound}
\end{align}
Simplifying Eq.~\eqref{eq:derivative_expression}, we get
\begin{align}
    \frac{\partial \varepsilon_{(2)}}{\partial \alpha} \bigg\rvert_{\alpha = \beta_c} &= \frac{4\beta_c(1-\beta_c^2)\Delta^4 + 4\beta_c\Delta^6}{(1-\beta_c^2)^6}  C_{6}^{(\mathrm{bound})}.
\end{align}
Substituting with the Taylor expansion for $C_{6}^{(\mathrm{bound})}$ from Eq.~\eqref{eq:Taylor_series_C6bound}, it follows that
\begin{align}
    &\frac{\partial \varepsilon_{(2)}}{\partial \alpha} \bigg\rvert_{\alpha = \beta_c}
    \approx \bigg( \frac{4\beta_c(1-\beta_c^2)\Delta^4 + 4\beta_c\Delta^6}{(1-\beta_c^2)^6}   \bigg) \nonumber \\
    &\times \bigg(1 + \frac{3\beta_c^2\Delta^2}{(1-\beta_c^2)^2} - \frac{3\beta_c^4\Delta^4}{(1-\beta_c^2)^4} + \frac{\beta_c^6\Delta^6}{(1-\beta_c^2)^6} \bigg), \nonumber \\
    &= \frac{4\beta_c}{(1-\beta_c^2)^5}\Delta^4 + O(\Delta^6),
\end{align}
yielding the result of Lemma~\ref{lemma:error_bound}.
\end{proof}

\end{appendices}

\bibliographystyle{IEEEtran}
\bibliography{IEEEabrv,ref}

\begin{thebibliography}{10}
\providecommand{\url}[1]{#1}
\csname url@samestyle\endcsname
\providecommand{\newblock}{\relax}
\providecommand{\bibinfo}[2]{#2}
\providecommand{\BIBentrySTDinterwordspacing}{\spaceskip=0pt\relax}
\providecommand{\BIBentryALTinterwordstretchfactor}{4}
\providecommand{\BIBentryALTinterwordspacing}{\spaceskip=\fontdimen2\font plus
\BIBentryALTinterwordstretchfactor\fontdimen3\font minus \fontdimen4\font\relax}
\providecommand{\BIBforeignlanguage}[2]{{%
\expandafter\ifx\csname l@#1\endcsname\relax
\typeout{** WARNING: IEEEtran.bst: No hyphenation pattern has been}%
\typeout{** loaded for the language `#1'. Using the pattern for}%
\typeout{** the default language instead.}%
\else
\language=\csname l@#1\endcsname
\fi
#2}}
\providecommand{\BIBdecl}{\relax}
\BIBdecl

\bibitem{Goodfellow-et-al-2016}
I.~Goodfellow, Y.~Bengio, and A.~Courville, \emph{Deep Learning}.\hskip 1em plus 0.5em minus 0.4em\relax MIT Press, 2016.

\bibitem{HeImageRecognition2016}
K.~{He}, X.~{Zhang}, S.~{Ren}, and J.~{Sun}, ``Deep {R}esidual {L}earning for {I}mage {R}ecognition,'' in \emph{2016 IEEE Conf. on Comput. Vis. Patt. Recog. (CVPR)}, June 2016, pp. 770--778.

\bibitem{GravesSpeechRecognition2013}
A.~{Graves}, A.~{Mohamed}, and G.~{Hinton}, ``Speech {R}ecognition with {D}eep {R}ecurrent {N}eural {N}etworks,'' in \emph{2013 IEEE Intl. Conf. Acoustics, Speech and Signal Process.}, May 2013, pp. 6645--6649.

\bibitem{bahdanau2014neural}
D.~Bahdanau, K.~Cho, and Y.~Bengio, ``{Neural Machine Translation by Jointly Learning to Align and Translate},'' in \emph{3rd Intl. Conf. on Learn. Rep., {ICLR} 2015, San Diego, CA, USA, May 7-9, 2015, Conference Track Proceedings}, 2015.

\bibitem{FUNAHASHI1993801}
K.~Funahashi and Y.~Nakamura, ``Approximation of dynamical systems by continuous time recurrent neural networks,'' \emph{Neur. Net.}, vol.~6, no.~6, pp. 801 -- 806, 1993.

\bibitem{AGUEROTORALES2021107373}
M.~M. Agüero-Torales, J.~I. {Abreu Salas}, and A.~G. López-Herrera, ``Deep learning and multilingual sentiment analysis on social media data: An overview,'' \emph{Applied Soft Computing}, vol. 107, p. 107373, 2021.

\bibitem{cho-etal-2014-learning}
K.~Cho, B.~van Merri{\"e}nboer, C.~Gulcehre, D.~Bahdanau, F.~Bougares, H.~Schwenk, and Y.~Bengio, ``{Learning Phrase Representations using {RNN} Encoder{--}Decoder for Statistical Machine Translation},'' in \emph{Proceedings of the 2014 Conference on Empirical Methods in Natural Language Processing ({EMNLP})}.\hskip 1em plus 0.5em minus 0.4em\relax Association for Computational Linguistics, Oct. 2014, pp. 1724--1734.

\bibitem{Guera2018}
D.~Güera and E.~J. Delp, ``{Deepfake Video Detection Using Recurrent Neural Networks},'' in \emph{2018 15th IEEE International Conference on Advanced Video and Signal Based Surveillance (AVSS)}, 2018, pp. 1--6.

\bibitem{Zhao2019}
B.~Zhao, X.~Li, and X.~Lu, ``{CAM-RNN: Co-Attention Model Based RNN for Video Captioning},'' \emph{IEEE Transactions on Image Processing}, vol.~28, no.~11, pp. 5552--5565, 2019.

\bibitem{mosleh2017brain}
S.~S. Mosleh, L.~Liu, C.~Sahin, Y.~R. Zheng, and Y.~Yi, ``Brain-{I}nspired {W}ireless {C}ommunications: {W}here {R}eservoir {C}omputing {M}eets {MIMO-OFDM},'' \emph{{IEEE} Trans. Neural Netw. Learn. Syst.}, vol.~29, no.~10, pp. 4694--4708, 2018.

\bibitem{peng2023rwkv}
B.~Peng, E.~Alcaide, Q.~Anthony, A.~Albalak, S.~Arcadinho, H.~Cao, X.~Cheng, M.~Chung, M.~Grella, K.~K. GV, X.~He, H.~Hou, P.~Kazienko, J.~Kocon, J.~Kong, B.~Koptyra, H.~Lau, K.~S.~I. Mantri, F.~Mom, A.~Saito, X.~Tang, B.~Wang, J.~S. Wind, S.~Wozniak, R.~Zhang, Z.~Zhang, Q.~Zhao, P.~Zhou, J.~Zhu, and R.-J. Zhu, ``{RWKV: Reinventing RNNs for the Transformer Era},'' 2023.

\bibitem{pascanu2013difficulty}
R.~Pascanu, T.~Mikolov, and Y.~Bengio, ``On the difficulty of training recurrent neural networks,'' in \emph{Intl. Conf. on Machine Learning}, 2013, pp. 1310--1318.

\bibitem{Werbos1990}
P.~Werbos, ``{Backpropagation Through Time: What It Does and How to Do It},'' \emph{Proceedings of the IEEE}, vol.~78, no.~10, pp. 1550--1560, 1990.

\bibitem{Hochreiter1997a}
S.~Hochreiter and J.~Schmidhuber, ``{{Long Short-Term Memory}},'' \emph{Neural Computation}, vol.~9, no.~8, pp. 1735--1780, 11 1997.

\bibitem{Bengio1994}
Y.~Bengio, P.~Simard, and P.~Frasconi, ``{Learning Long-Term Dependencies with Gradient Descent is Difficult},'' \emph{{IEEE} Trans. Neural Netw.}, vol.~5, no.~2, pp. 157--166, 1994.

\bibitem{Hochreiter1996}
S.~Hochreiter and J.~Schmidhuber, ``{LSTM can Solve Hard Long Time Lag Problems},'' in \emph{Advances in Neural Information Processing Systems}, M.~Mozer, M.~Jordan, and T.~Petsche, Eds., vol.~9.\hskip 1em plus 0.5em minus 0.4em\relax MIT Press, 1996.

\bibitem{SHERSTINSKY2020}
A.~Sherstinsky, ``{Fundamentals of Recurrent Neural Network (RNN) and Long Short-Term Memory (LSTM) network},'' \emph{Physica D: Nonlinear Phenomena}, vol. 404, p. 132306, 2020.

\bibitem{Gallicchio2018RandomizedRN}
C.~Gallicchio, A.~Micheli, and P.~Tiňo, ``{Randomized Recurrent Neural Networks},'' in \emph{The European Symposium on Artificial Neural Networks}, 2018.

\bibitem{lukovsevivcius2009reservoir}
M.~Luko{\v{s}}evi{\v{c}}ius and H.~Jaeger, ``Reservoir computing approaches to recurrent neural network training,'' \emph{Computer Science Review}, vol.~3, no.~3, pp. 127--149, 2009.

\bibitem{Lukosevicius2012}
M.~Luko{\v{s}}evi{\v{c}}ius, \emph{A Practical Guide to Applying Echo State Networks}.\hskip 1em plus 0.5em minus 0.4em\relax Springer Berlin Heidelberg, 2012, pp. 659--686.

\bibitem{Hinaut2012}
X.~Hinaut and P.~F. Dominey, ``{On-Line Processing of Grammatical Structure Using Reservoir Computing},'' in \emph{Artificial Neural Networks and Machine Learning -- ICANN 2012}, A.~E.~P. Villa, W.~Duch, P.~{\'E}rdi, F.~Masulli, and G.~Palm, Eds.\hskip 1em plus 0.5em minus 0.4em\relax Berlin, Heidelberg: Springer Berlin Heidelberg, 2012, pp. 596--603.

\bibitem{Juven2020}
A.~Juven and X.~Hinaut, ``{Cross-Situational Learning with Reservoir Computing for Language Acquisition Modelling},'' in \emph{2020 International Joint Conference on Neural Networks (IJCNN)}, 2020, pp. 1--8.

\bibitem{Jalalvand2015}
A.~Jalalvand, G.~Van~Wallendael, and R.~Van De~Walle, ``{Real-Time Reservoir Computing Network-Based Systems for Detection Tasks on Visual Contents},'' in \emph{2015 7th Intl Conf. on Computational Intelligence, Communication Systems and Networks}, 2015, pp. 146--151.

\bibitem{WANG2021115022}
W.-J. Wang, Y.~Tang, J.~Xiong, and Y.-C. Zhang, ``{Stock market index prediction based on reservoir computing models},'' \emph{Expert Systems with Applications}, vol. 178, p. 115022, 2021.

\bibitem{zhou2019}
Z.~Zhou, L.~Liu, and H.-H. Chang, ``Learning for {D}etection: {MIMO-OFDM} {S}ymbol {D}etection {T}hrough {D}ownlink {P}ilots,'' \emph{{IEEE} Trans. Wireless Commun.}, vol.~19, no.~6, pp. 3712--3726, 2020.

\bibitem{zhou2020rcnet}
Z.~Zhou, L.~Liu, S.~Jere, J.~Zhang, and Y.~Yi, ``{RCNet}: {I}ncorporating {S}tructural {I}nformation {I}nto {D}eep {RNN} for {O}nline {MIMO-OFDM} {S}ymbol {D}etection {W}ith {L}imited {T}raining,'' \emph{{IEEE} Trans. Wireless Commun.}, vol.~20, no.~6, pp. 3524--3537, 2021.

\bibitem{RCStruct_TWC}
J.~Xu, Z.~Zhou, L.~Li, L.~Zheng, and L.~Liu, ``{RC}-{S}truct: {A} {S}tructure-based {N}eural {N}etwork {A}pproach for {MIMO-OFDM} {D}etection,'' \emph{{IEEE} Trans. Wireless Commun.}, pp. 1--1, 2022.

\bibitem{Ren2020}
H.-P. Ren, H.-P. Yin, C.~Bai, and J.-L. Yao, ``{Performance Improvement of Chaotic Baseband Wireless Communication Using Echo State Network},'' \emph{{IEEE} Trans. Commun.}, vol.~68, no.~10, pp. 6525--6536, 2020.

\bibitem{HaoHsuanDEQN2022}
H.-H. Chang, L.~Liu, and Y.~Yi, ``{Deep Echo State Q-Network (DEQN) and Its Application in Dynamic Spectrum Sharing for 5G and Beyond},'' \emph{{IEEE} Trans. Neural Netw. Learn. Syst.}, vol.~33, no.~3, pp. 929--939, 2022.

\bibitem{DaRos2020}
F.~Da~Ros, S.~M. Ranzini, H.~Bülow, and D.~Zibar, ``{Reservoir-Computing Based Equalization With Optical Pre-Processing for Short-Reach Optical Transmission},'' \emph{IEEE Journal of Selected Topics in Quantum Electronics}, vol.~26, no.~5, pp. 1--12, 2020.

\bibitem{Dai2021}
H.~Dai and Y.~K. Chembo, ``{Classification of IQ-Modulated Signals Based on Reservoir Computing With Narrowband Optoelectronic Oscillators},'' \emph{IEEE Journal of Quantum Electronics}, vol.~57, no.~3, pp. 1--8, 2021.

\bibitem{Rubayet20AI}
R.~Shafin, L.~Liu, V.~Chandrasekhar, H.~Chen, J.~Reed, and J.~C. Zhang, ``{Artificial Intelligence-Enabled Cellular Networks: A Critical Path to Beyond-5G and 6G},'' \emph{{IEEE} Wireless Commun.}, vol.~27, no.~2, pp. 212--217, 2020.

\bibitem{Montavon2019}
G.~Montavon, A.~Binder, S.~Lapuschkin, W.~Samek, and K.-R. M{\"u}ller, \emph{Layer-Wise Relevance Propagation: An Overview}.\hskip 1em plus 0.5em minus 0.4em\relax Cham: Springer International Publishing, 2019, pp. 193--209.

\bibitem{shwartzziv2017opening}
R.~Shwartz-Ziv and N.~Tishby, ``{Opening the Black Box of Deep Neural Networks via Information},'' 2017.

\bibitem{Lundberg2017}
S.~M. Lundberg and S.-I. Lee, ``{A Unified Approach to Interpreting Model Predictions},'' in \emph{Proceedings of the 31st Intl Conf. on Neural Information Processing Systems}, ser. NIPS'17.\hskip 1em plus 0.5em minus 0.4em\relax Red Hook, NY, USA: Curran Associates Inc., 2017, p. 4768–4777.

\bibitem{Baehrens2010}
D.~Baehrens, T.~Schroeter, S.~Harmeling, M.~Kawanabe, K.~Hansen, and K.-R. M\"{u}ller, ``How to explain individual classification decisions,'' \emph{J. Mach. Learn. Res.}, vol.~11, p. 1803–1831, aug 2010.

\bibitem{Ozturk2007}
M.~C. Ozturk, D.~Xu, and J.~C. Pr\'{\i}ncipe, ``{Analysis and Design of Echo State Networks},'' \emph{Neural Comput.}, vol.~19, no.~1, p. 111–138, 2007.

\bibitem{Bollt2021}
E.~Bollt, ``On explaining the surprising success of reservoir computing forecaster of chaos? {T}he universal machine learning dynamical system with contrast to {VAR} and {DMD},'' \emph{Chaos}, vol.~31, p. 013108, 2021.

\bibitem{HART2021}
A.~G. Hart, J.~L. Hook, and J.~H. Dawes, ``Echo {S}tate {N}etworks trained by {T}ikhonov least squares are {L}2($\mu$) approximators of ergodic dynamical systems,'' \emph{Physica D: Nonlinear Phenomena}, vol. 421, p. 132882, 2021.

\bibitem{Halus2019}
A.~Haluszczynski and C.~Räth, ``Good and bad predictions: Assessing and improving the replication of chaotic attractors by means of reservoir computing,'' \emph{Chaos}, vol.~29, no.~10, p. 103143, 2019.

\bibitem{Carroll2022}
T.~L. Carroll, ``Optimizing memory in reservoir computers,'' \emph{Chaos}, vol.~32, no.~2, p. 023123, 2022.

\bibitem{JereTCOM2023}
S.~Jere, R.~Safavinejad, and L.~Liu, ``{Theoretical Foundation and Design Guideline for Reservoir Computing-based MIMO-OFDM Symbol Detection},'' \emph{{IEEE} Trans. Commun.}, pp. 1--1, 2023.

\bibitem{Gonon2020}
L.~Gonon, L.~Grigoryeva, and J.-P. Ortega, ``{R}isk {B}ounds for {R}eservoir {C}omputing,'' \emph{Jour. Mach. Learn. Res.}, vol.~21, no. 240, pp. 1--61, 2020.

\bibitem{Jere2023WCL}
S.~Jere, R.~Safavinejad, L.~Zheng, and L.~Liu, ``{Channel Equalization Through Reservoir Computing: A Theoretical Perspective},'' \emph{{IEEE} Wireless Commun. Lett.}, vol.~12, no.~5, pp. 774--778, 2023.

\bibitem{Oppenheim1996}
A.~V. Oppenheim, A.~S. Willsky, and S.~H. Nawab, \emph{Signals \& Systems (2nd ed.)}.\hskip 1em plus 0.5em minus 0.4em\relax USA: Prentice-Hall, Inc., 1996.

\bibitem{Sozos2022}
K.~Sozos, A.~Bogris, P.~Bienstman, G.~Sarantoglou, S.~Deligiannidis, and C.~Mesaritakis, ``High-speed photonic neuromorphic computing using recurrent optical spectrum slicing neural networks,'' \emph{Communications Engineering}, vol.~1, no.~1, p.~24, 10 2022.

\bibitem{LIU2008998}
X.~Liu, ``A new method for the pole estimation of linear time-invariant systems using singular value decomposition,'' \emph{Journal of Sound and Vibration}, vol. 310, no.~4, pp. 998--1013, 2008.

\bibitem{ZHANG2005367}
Y.~Zhang, Z.~Zhang, X.~Xu, and H.~Hua, ``Modal parameter identification using response data only,'' \emph{Journal of Sound and Vibration}, vol. 282, no.~1, pp. 367--380, 2005.

\bibitem{edelman_rao_2005}
A.~Edelman and N.~R. Rao, ``Random matrix theory,'' \emph{Acta Numerica}, vol.~14, p. 233–297, 2005.

\bibitem{Bianchi2018}
F.~M. Bianchi, L.~Livi, and C.~Alippi, ``{Investigating Echo-State Networks Dynamics by Means of Recurrence Analysis},'' \emph{{IEEE} Trans. Neural Netw. Learn. Syst.}, vol.~29, no.~2, pp. 427--439, 2018.

\bibitem{Brunton_Kutz_2019}
S.~L. Brunton and J.~N. Kutz, \emph{Data-Driven Science and Engineering: Machine Learning, Dynamical Systems, and Control}.\hskip 1em plus 0.5em minus 0.4em\relax Cambridge University Press, 2019.

\bibitem{Akaike1998}
H.~Akaike, \emph{Information Theory and an Extension of the Maximum Likelihood Principle}.\hskip 1em plus 0.5em minus 0.4em\relax New York, NY: Springer New York, 1998, pp. 199--213.

\bibitem{Stoica2004}
P.~Stoica and Y.~Selen, ``{Model-Order Selection: A review of information criterion rules},'' \emph{{IEEE} Signal Process. Mag.}, vol.~21, no.~4, pp. 36--47, 2004.

\bibitem{robert1999monte}
C.~P. Robert and G.~Casella, \emph{{Monte Carlo Statistical Methods}}.\hskip 1em plus 0.5em minus 0.4em\relax Springer, 1999, vol.~2.

\bibitem{Caffo2002}
B.~S. Caffo, J.~G. Booth, and A.~C. Davison, ``{Empirical supremum rejection sampling},'' \emph{Biometrika}, vol.~89, no.~4, pp. 745--754, 12 2002.

\end{thebibliography}

\end{document}